\documentclass[journal = jctcce,manuscript=article]{achemso}
\setkeys{acs}{usetitle = true}

\usepackage{amsmath}
\usepackage{amssymb}
\usepackage{graphicx}
\usepackage[colorlinks=true,citecolor=blue,linkcolor=blue]{hyperref}

\newcommand{\Cov}[1] {\mathrm{cov}\left( #1 \right)}
\newcommand{\Var}[1] {\mathrm{var}\left( #1 \right)}

\title[Simple quantitative tests to validate]{Simple quantitative tests to validate sampling from thermodynamic ensembles}

\author{Michael R. Shirts}
\email{michael.shirts@virginia.edu}
\affiliation{Department of Chemical Engineering, University of Virginia, VA 22904}

\date{\today}

\begin{document}


\begin{abstract}
It is often difficult to quantitatively determine if a new molecular
simulation algorithm or software properly implements sampling of the
desired thermodynamic ensemble. We present some simple statistical
analysis procedures to allow sensitive determination of whether a
desired thermodynamic ensemble is properly sampled.  We demonstrate
the utility of these tests for model systems and for molecular
dynamics simulations in a range of situations, including constant
volume and constant pressure simulations, and describe an
implementation of the tests designed for end users.
\end{abstract}


\section{Introduction}
\label{section:introduction}

Molecular simulations, including both molecular dynamics (MD) and
Monte Carlo (MC) techniques, are powerful tools to study the
properties of complex molecular systems.  When used to specifically
study thermodynamics of such systems, rather than dynamics, the
primary goal of molecular simulation is to generate uncorrelated
samples from the appropriate ensemble as efficiently as possible.
This simulation data can then be used to compute thermodynamic
properties of interest. Simulations of several different ensembles may
be required to simulate some thermodynamic properties, such as free
energy differences between states. An ever-expanding number of
techniques have been proposed to perform more and more sophisticated
sampling from complex molecular systems using both MD and MC, and new
software tools are continually being introduced in order to implement
these algorithms and to take advantage of advances in hardware
architecture and programming languages.

However, it is extremely easy to make subtle errors in both the
theoretical development and the computer implementation of these
advanced sampling algorithms.  Such errors can occur because of
numerical errors in the underlying energy functions, theoretical
errors in the proposed algorithm, approximations that are too extreme,
and the programming bugs that are inevitable when managing more and
more complicated code bases.

There are a number of reasons it is difficult to validate a given
implementation of an algorithm for the proper thermodynamic behavior.
First, we lack analytical results for virtually all complex molecular
systems, and analytically soluble toy problems may not have all of the
features that more complicated systems of actual research interest may
possess.  Additionally, molecular simulations generate statistical
samples from the probability distribution of the system.  Most
observables therefore require significant simulation time to reduce
statistical noise to a level sufficiently low to allow conclusive
identification of small but potentially significant violations of the
sampled ensembles.

There are of course some aspects of molecular distributions that can
and should always be checked directly.  For example, in an NVE
ensemble the total energy should be conserved with statistically zero
drift.  For symplectic integrators with NVE simulations, the RMS error
will scale with the square of the step size.  For an NVT ensemble when
the potential energy is independent of particle momenta (which is true
with the rare exception of systems with magnetic forces), the kinetic
energy will follow the Maxwell-Boltzmann distribution and the
consistency of sampled results can be tested against this distribution
with standard statistical methods.  NVT simulations must have an
average kinetic energy corresponding to the desired temperature, and
NPT simulations must have the proper average instantaneous pressure
computed from the virial and kinetic energy. However, there are no
standard tests for proper distribution for the potential energy, which
greatly complicates Monte Carlo simulations, or for total energy of an
arbitrary simulation system. Additionally, there are many possible
distributions which have the correct average temperature or pressure,
but do not satisfy the proper Boltzmann probability distributions for
our specific ensemble of interest.

It is therefore worthwhile to have physically rigorous strategies and
tools for assessing whether a simulation method is indeed generating
samples from the desired distribution in its entirety.  Such general
strategies could help to better answer vital questions such as ``Is
this thermostat/barostat correct?,'' ``How much does a very long time
step affect my energy distribution?'' and of course, ``Have I {\em
  finally} got all the bugs out of my code now?''

\section{Theory}

Thermodynamic ensembles all have similar probability distributions
with respect to macroscopic intensive parameters and microstates,
e.g.:
\begin{eqnarray}
P(\vec{x}|\beta) \propto \exp(-\beta H(\vec{p},\vec{q})) &\hspace{1 cm} & \text{canonical} \label{eq:canonical}\\
P(\vec{x},V|\beta,P) \propto \exp(-\beta (H(\vec{p},\vec{q}) + PV)) & \hspace{1 cm}&  \text{isobaric-isothermal}  \label{eq:isobaric}\\
P(\vec{x},\vec{N}|\beta,\vec{\mu}) \propto \exp(-\beta (H(\vec{p},\vec{q}) - \sum_{\mathrm{species}} \mu_i N_i)) & \hspace{1 cm}& \text{grand canonical}  \label{eq:grandcanonical}
\end{eqnarray}
where $P(a|b)$ indicates the probability of a microstate determined by
variable or variables $a$ given macroscopic parameter or parameters
$b$. Specifically, all have the exponential form $\exp(-u(\vec{x}))$
where $\vec{x} = (\vec{p},\vec{q},V,\vec{N})$ is the microstate and
$u(\vec{x})$ is a reduced energy term whose form depends on the
ensemble.

This reduced energy term is a generalized function of two types of
variables.  The first type of variables are the degrees of freedom
determining the microstates of each ensemble, including the positions
and velocities of the atoms, but also potentially including the volume
of the system $V$ and the number of particles of each of $i$ species
in the system $N_i$.  The second type of variables are those
determining the ensemble of the physical system, including the
temperature $T$, the pressure $P$, the chemical potentials $\mu_i$,
and the specific functional form of the Hamiltonian $H(\vec{p},\vec{q})$.  These
equations, along with the requirement that all microstates with the
same value for the generalized energy term have the same probability,
completely define the thermodynamic ensemble.  A general test should
therefore check as directly as possible that the samples we collect
are fully consistent with
Eqs.~\ref{eq:canonical}--\ref{eq:grandcanonical}.  For simplicity, we
will perform an initial derivation of such a test using the canonical
ensemble, and then generalize the derivation to other ensembles.

The probability density of observing a specific energy in the
canonical ensemble (Eq.\ref{eq:canonical}) can be written in terms of
the density of states $\Omega(E) = \exp(S(N,V,E)/k_B)$ as
\begin{equation}
 P(E|\beta) = Q(\beta)^{-1} \Omega(E) \exp(-\beta E)
\end{equation}
where $S$ is the entropy, $\beta = (k_B T)^{-1}$, $k_B$ is Boltzmann's
constant, and $Q(\beta) = \int \Omega(E) \exp(-\beta E) dE $ is the
canonical partition function, related to the Helmholtz free energy $A$
by $A = -\beta^{-1} \ln Q$. $Q$ is a function of $\beta$, but not $E$,
whereas $\Omega$ is a function of $E$, but importantly, not
$\beta$. Note that at this point, $E$ is specifically the total
energy, though we will examine kinetic and potential energies
separately later on.

Without specific knowledge of what the density of states $\Omega(E)$
is for a particular molecular system, no quantity of samples from a
single state can identify if the energies indeed have the proper
distribution. However, if we take the ratio of the probability
distributions of two simulations performed at different temperatures,
hence with two different values of $\beta$, but with otherwise
identical parameters, the unknown density of states cancels leaving:
\begin{eqnarray}
\frac{P(E|\beta_2)}{P(E|\beta_1)} &=& \frac{\frac{\exp(-\beta_2 E)}{Q(\beta_2)}}{\frac{\exp(-\beta_1 E)}{Q(\beta_1)}} \nonumber \\
                                  &=& \exp([\beta_2 A_2 -\beta_1 A_1] -[\beta_2 -\beta_1] E)
\end{eqnarray}
If we take the logarithm of this ratio, we obtain:
\begin{eqnarray}
\ln \frac{P(E|\beta_2)}{P(E|\beta_1)} &=& [\beta_2 A_2 -\beta_1 A_1] -[\beta_2 -\beta_1] E 
\label{eq:canonicaltest}
\end{eqnarray}
which is of the linear form $\alpha_0 + \alpha_1 E$. Note that linear
coefficient $\alpha_1 = -(\beta_2-\beta_1)$ is independent of the
(unknown in general) Helmholtz free energies $A_2$ and $A_1$.  

This relationship forms the basis of the ensemble validation
techniques we present in this paper. Similar formulas can be derived
for any of the standard thermodynamic ensembles with probability
distributions of the form $e^{-u(\vec{x})}$ as long as the reduced
energy term is linear in conjugate parameters.  Non-exponential
probability distributions are certainly possible to generate in
simulations, but are much less standard, and so we will not deal
directly with them in this study.  The same general techniques will
work if the probability of a given microstate depends only on the
energy of the microstate.  We will call agreement of a simulation with
its target distribution as described by Eq.~\ref{eq:canonicaltest} and
its analogs for other ensembles as {\em ensemble consistency}.

There are a number of ways to check if the distribution of samples from a
given pair of simulations satisfies these equations.  The most
straightforward way starts with binning the energies $E$ from both
simulations. If the distributions are sufficiently close together to
have statistically well-defined probabilities at overlapping values of
$E$ and we have sufficient data, we can fit the ratio of the
histogram probabilities to a line in this overlap region.  If the
slope deviates from $-(\beta_2-\beta_1)$ by a statistically
significant amount, then the data necessarily deviates from a
canonical distribution. However, deciding quantitatively what
constitutes ``statistically significant'' can be challenging, and will
be further explored in this paper.

This test of consistency with Eq.~\ref{eq:canonicaltest} is a {\em
  necessary} test for an algorithm that is consistent with the
canonical ensemble; if the slope of the probability ratio deviates
from the true line, the data cannot be consistent with the
ensemble. However, the test is not necessarily a {\em sufficient} test
of simulation quality as it does not include any direct test of
ergodicity.  Specifically, it says nothing about whether states with
the same energy are sampled with equal probability as is required by
statistical mechanics. It also does not say anything about whether
there are states that are not sampled.  We could have sampling
consistent with the desired ensemble but trapped in only a small
portion of the allowed phase space of a system.

In general, additional tests of convergence or ergodicity are required
before the system can be guaranteed to be sampled correctly.  For
example, for molecular dynamics, one could examine the kinetic energy
of different partitions of the degrees of freedom as can be used to
diagnose such problems as the ``flying ice cube,'' occurring in some
poorly configured simulations when the center of mass degrees of
freedom are decoupled from other degrees of
freedom.~\cite{icecube_1998} However, for testing algorithms or code,
simple systems that are both sufficiently complicated and general can
usually be found which will behave ergodically within a reasonable
amount of simulation time.  Therefore, in the rest of this paper, we
will analyze systems which are clearly sampled ergodically and which
have converged ensemble averages of interest, so we will not require
any additional tests of ergodicity or convergence.

Having analyzed the potential problems with such ensemble validation
analysis, we next explore possible methods to quantify deviation
from the canonical ensemble using data collected from pairs of
simulations.

\subsection{Visual Inspection}

\begin{figure}[tbp]
\noindent
\begin{tabular}{cc}
\includegraphics[width=0.5\columnwidth]{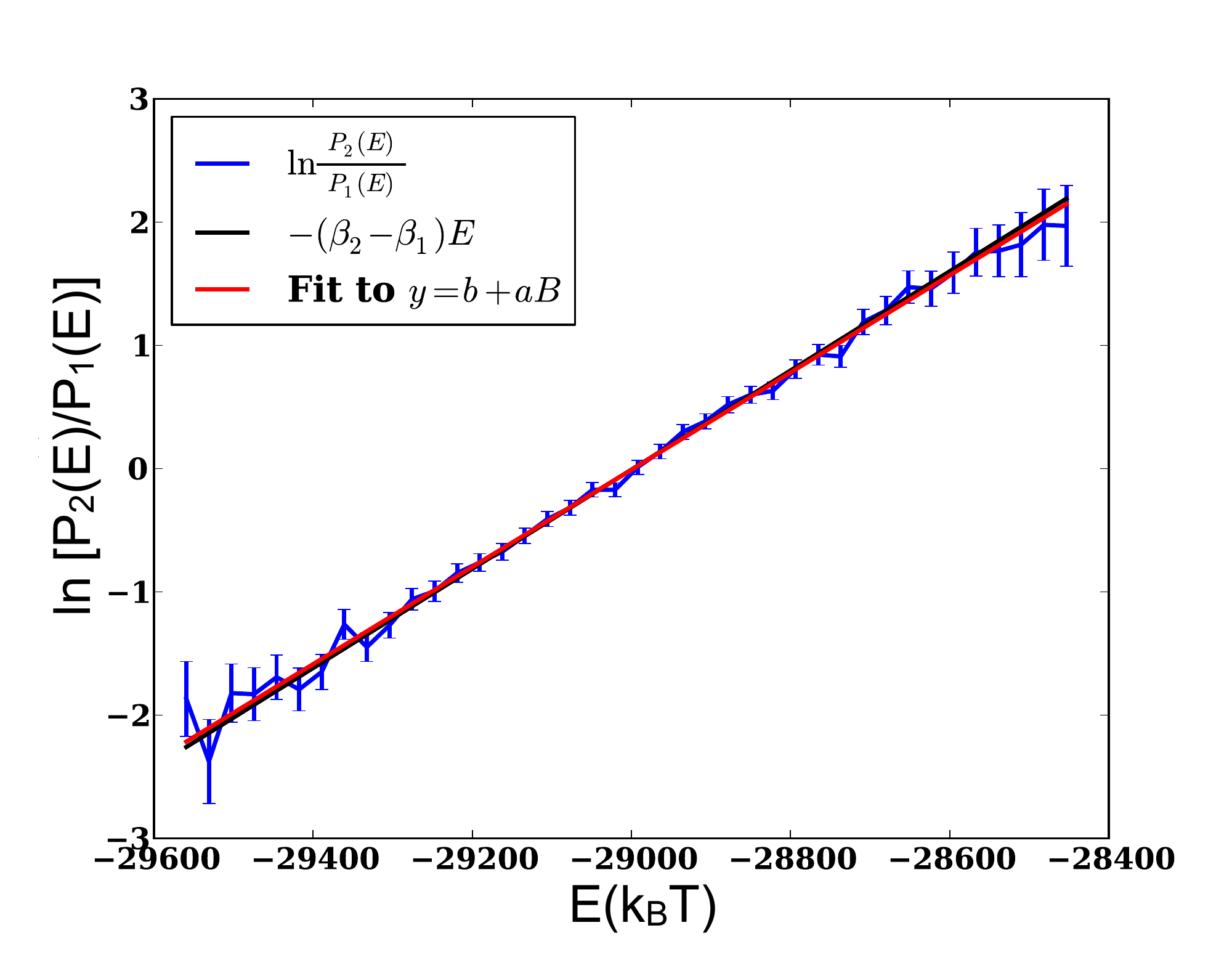} & \includegraphics[width=0.5\columnwidth]{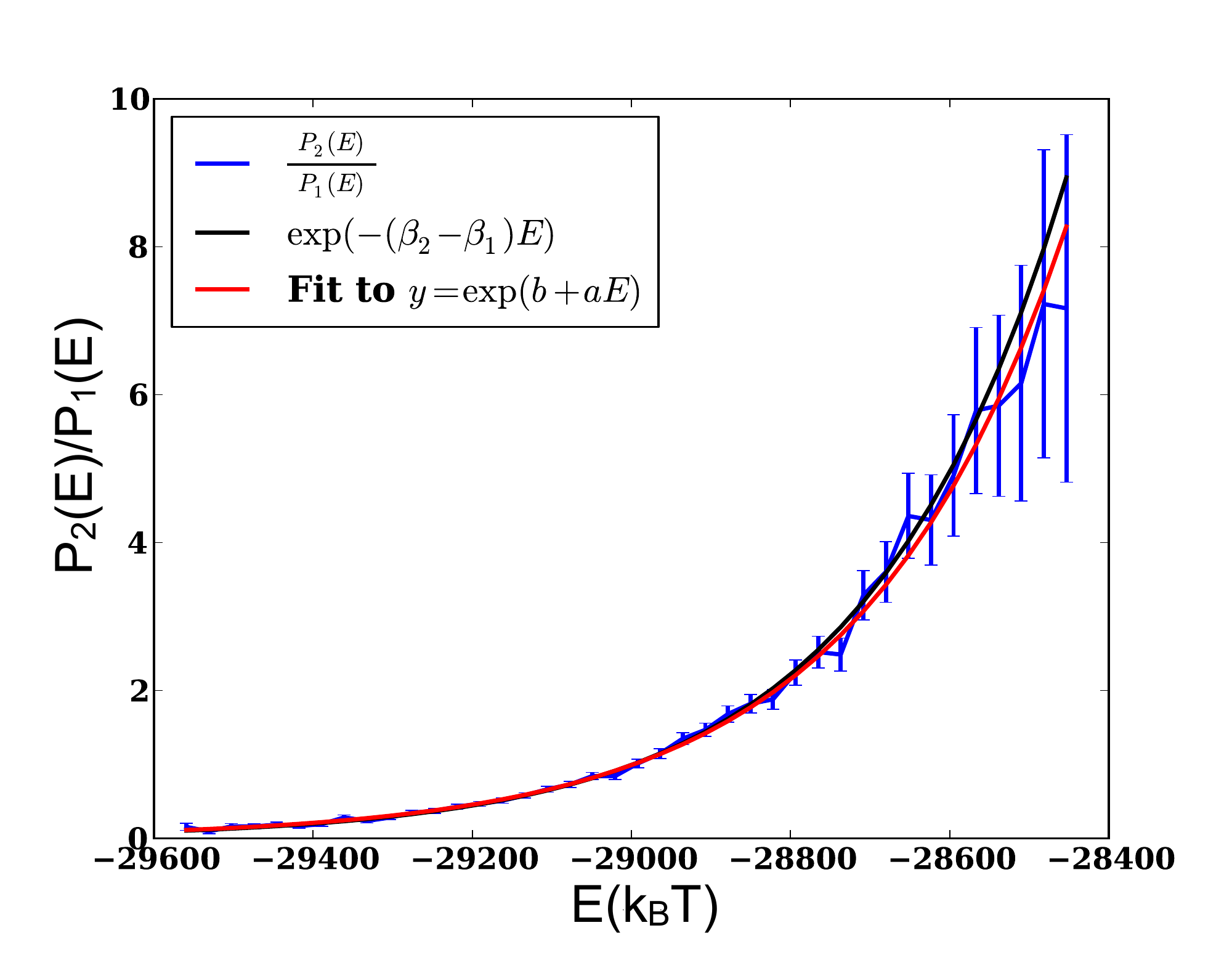} \\
a  &  b 
\end{tabular}
\caption{{\bf Ensemble validation of water simulations.} Validation of
  the energy distribution of 900 TIP3P water molecules simulated in
  the NVT ensemble using the Nos\'{e}-Hoover
  algorithm~\protect{\cite{Hoover1985}}.  The predicted value and the
  actual data for both linear (a) and nonlinear (b)
  fits quantitatively agree.\label{fig:example_plot}}
\end{figure}

We can divide the common energy range of the two simulations into bins
(perhaps 20-40, depending on the amount of data, numbers chosen solely
from experience through trial and error).  Bins need not be equally
spaced, though this simplifies the analysis considerably by removing
the need to correct the probability densities for differing width of
bins.  It is also greatly simplifies the analysis to select bin
divisions that are aligned between the two data sets.  Bins can be
chosen to exclude a few points on top and the bottom of each
distribution to avoid small sample error and zero densities at the
extremes. $P_1(E)$ and $P_2(E)$ in each bin can then be estimated
directly from the histograms. We can compute the ratio of these
histograms at each value of the energy at the centers of the bins, and
plot either the ratio, or more cleanly, logarithm of this ratio, as
shown in Fig.~\ref{fig:example_plot}. If this logarithm ratio is
linear, we have a system that for all {\em qualitative} purposes obeys
the proper equilibrium distribution.

Qualitatively, if the actual slope of the log energy ratios is below
the expected, slope, it means that the low $\beta$ (high temperature)
simulation samples that particular energy less than it should, while
if it is above the true line, it means that portion of the
distribution is oversampled.  A consistently higher observed slope
therefore means that the distribution is narrower than it should be,
and a lower observed slope means that the distribution is wider than
it should be.

\subsection{Quantitative Fitting}
The relationships presented so far are not entirely novel; visual
inspection of probability ratios of paired temperature replica
exchange simulations has been used previously to check that
neighboring replicas have the proper distributions relative to each
other.~\cite{Rhee2003, Sindhikara2008} However, there does not appear
to have been an effort to use this relationship as a general test to
quantitatively analyze simulations for goodness-of-fit to the putative
ensemble distributions.

\subsubsection{Linear fitting} To make this ensemble test quantitative, we estimated the error in
the occupancy probability $p_k$ of each bin $i$ as $\delta p_k =
\sqrt{p_k(1-p_k)/n}$ (a standard probability result), and propagate
the error in the individual bin probabilities into the ratio
$P_2(E)/P_1(E))$ (a process detailed in
Appendix~\ref{appendix:error}). If the true slope lies consistently
outside of the error estimates then it is very likely the simulation
is not correctly sampling the desired ensemble. Calculation of the
histogram uncertainties also allows us to perform weighted linear and
nonlinear least squares fitting (details also in
Appendix~\ref{appendix:error}).  This allows us to includes the effect
of small sample error at the extremes of the distribution in our
fitting.  We can use standard error propagation methods to propagate
the error in the histogram occupancy ratios into the error in the
linear parameters.

\subsubsection{Nonlinear fitting}
 It is well-known that linearizing a mathematical relationship in
 order to perform linear least squares can introduce bias in the
 estimation of the parameters.  It is therefore often preferable to
 minimize the direct sum of the residuals $S_r(\vec{\alpha}) = \sum_{i}
 (y_i - f(\vec{\alpha},x_i))^2$, which is a nonlinear function of
 $\vec{a}$, and then propagate the error in the histogram bins into
 the uncertainties of the components of $\vec{\alpha}$.  In this particular
 problem, we want to determine the two parameter fit that minimizes
 the sum of residuals $S_r(\alpha_0,\alpha_1)$ for the function
\begin{eqnarray}
S_r(\alpha_0,\alpha_1) = \sum_{i} \left[\frac{P_1(E_i)}{P_2(E_i)} - \exp(\alpha_0+\alpha_1 E_i)\right]^2
\end{eqnarray}

\subsubsection{Maximum likelihood estimates}
Any results from either the linear or nonlinear case may affected by
the choice of histogram binning we use.  In theory, we can vary the
number of histogram bins to ensure the answers are not dependent on
the number of bins. However, we can completely eliminate the histogram
dependence as well as including the data at the tails rather than
truncating them by using a maximum likelihood approach. A maximum
likelihood approach allows us to predict the most likely parameters
for a given statistical model from the data that has been observed.

Previously, we have used such a maximum likelihood approach to compute
the free energy difference between forward and reverse work distributions
between two thermodynamic states at the same
temperature,~\cite{Shirts2003} which is equivalent to computing the
value of $\beta^{-1}\ln Q_1/Q_2$ with fixed $\beta$.  In the present
case, we now have {\em two} parameters in the distribution which we
must fit, $\alpha_0 = \ln Q_1/Q_2$ and $\alpha_1 =
-(\beta_2-\beta_1)$.  Applying a maximum likelihood approach along the
lines described in the paper~\cite{Shirts2003} leads to log likelihood
equations:
\begin{eqnarray*} 
\ln L (\alpha|\text{data}) &=& \sum_{i=1}^{N_1} \ln f(-\alpha_0 - \alpha_1 E_i) + \sum_{j=1}^{N_2} \ln f(\alpha_0 + \alpha_1 E_j)
\label{eq:maxlike}
\end{eqnarray*}
where $f(x)$ is the Fermi function $f(x) = [1+\exp(-x)]^{-1}$, and where the
first sum is over energies sampled at temperature $T_1$ and the second
sum is over the energies sampled at $T_2$.  The most likely
parameters are the ones which maximize the likelihood function in
Eq.~\ref{eq:maxlike}.  This particular function can be shown to have
no minima and only one maximum, so it will always converge.

Eq.~\ref{eq:maxlike} can be solved by any of the standard techniques
for multidimensional optimization as it is everywhere concave.  There
is one minor technicality; clearly, the variance can be minimized to
zero by setting $\alpha_0=\alpha_1=0$, which is not physically
consistent with the data.  There is therefore an additional constraint
we must first identify to find a unique minimum.

In performing this likelihood maximization, we note that although
there are four parameters explicitly stated, $A_1$, $A_2$, $\beta_1$
and $\beta_2$, only two of them are actually free parameters.
Examining Eq.~\ref{eq:canonicaltest}, we can express the relationship
to the physical quantities as $\alpha_0 = \beta_2 A_2 - \beta_1 A_1$
and $\alpha_1 = -(\beta_2-\beta_1)$.  We also note that
Eq.~\ref{eq:canonicaltest} does not allow us to test for $\beta_1$ and
$\beta_2$ directly, but instead is only a function of the difference
$\beta_2-\beta_1$, so we must actually treat this as one variable
corresponding to a single degree of freedom.  A simple choice is to
treat $\beta_1 + \beta_2$ as a constant in what amounts to a choice of
the energy scale.  We can therefore set $\beta_{\mathrm{ave}} =
\frac{1}{2}(\beta_{1,user} + \beta_{2,user})$, the user specified
temperatures. $A_1$ and $A_2$ are the free energies of the system, so
there is no physical meaning to their absolute value, only their
difference.  Without loss of generality, we set $A_1 + A_0 = 0$, and
treat $\Delta A = A_2 - A_1$ as our second independent variable.
These two choices allow us to then solve for unique values of
$\alpha_0$ and $\alpha_1$, rewriting $\alpha_0 + \alpha_1 E =
\beta_{\text{ave}}(A_2 - A_1) -(\beta_2-\beta_1)E =
\beta_{\text{ave}}\Delta A - \Delta \beta E$, an expression that
explicitly only has two free parameters.

One downside of using a maximum likelihood analysis is that it does
not give a graphical representation; it is histogram independent, and
so we do not have a histogram that we can plot!  A linear fit should
therefore be performed in conjunction with maximum likelihood analysis
to quickly visualize the data as a sanity check.

\subsection{Error estimates}

Once we have an estimate of the slope $\beta_2 - \beta_1$, we must ask
if the slope deviates from the true value with a statistically
significant deviation or if the difference more likely due to
statistical variation.  For this, we can turn to error estimation
techniques to find a statistically robust approximation for the error
in $\beta_2 - \beta_1$ and to determine if any deviations from the
true value are most likely a result of statistical noise or actual
errors in the simulation.

For weighted linear least squares, weighted nonlinear least squares,
and multiparameter maximum likelihood logistic regression, the
analytic asymptotic error estimators for the covariance matrix of
fitting parameters are all well-known statistical results:
\begin{eqnarray*}
\text{linear}  &\hspace{1 cm} \Cov{\vec{\alpha}} = &(X^T WX)^{-1} \\
\text{nonlinear} &\hspace{1 cm} \Cov{\vec{\alpha}} =& (J^T W^{-1} J)^{-1}\\
\text{maximum likelihood} &\hspace{1 cm} \Cov{\vec{\alpha}} =& (\mathrm{Hess}(\ln L)_{{\bf \alpha}})^{-1}
\end{eqnarray*}
In all equations $\vec{\alpha}$ is the vector of parameters we are
estimating.  In the first equation, $X$ is the $(M+1)\times N$ matrix
with the first column all ones, and the second through $(M+1)$th
column the values of the $N$ observations of the $M$ observables.  In
the second equation $J$ is the Jacobian of the model with respect to
the vector of parameters, evaluated at $N$ observations and the values
of the parameters minimizing the nonlinear fit. In the last equation,
$\mathrm{Hess}(\ln L)$ is the Hessian of log likelihood with respect
to the parameters, and $W$ is a weight matrix consisting of the
variances of the values of each data point estimated from the
histograms.  We explore these expressions more completely in the
Appendices~\ref{appendix:error} and~\ref{appendix:mle}.

We can also use bootstrap sampling of the original distribution data
to generate error estimates, which has proven to be a reliable error
estimation method for free energy calculations.~\cite{Shirts2011}
Although more computationally intensive, the total burden is
relatively low. For example, it takes only 20 minutes on a single core of
a 2.7 GHz Intel i7 processor to perform 200 bootstrap samples, even
with 600 000 energy evaluations from each simulation.

Once we have generated error estimates for our estimates of the
parameters, we can ask the underlying statistical question of whether
deviations from the true result are likely caused by statistical error
or by errors in the underlying data.  In most cases we will have
collected enough samples that the deviation from the fit should be
distributed normally.  In this case, we can simply compute the
standard deviation of the fit parameters and ask how many standard
deviations the calculated slope $\beta_2-\beta_1$ is from the user
specified slope. If this difference is consistently more than 2-3
$\sigma$ away from the true value in repeated tests, it indicates that
there are likely errors with the simulations as the two distributions
do not have the relationship that they would have if they obeyed a
canonical distribution.  More sophisticated statistical tests are
possible that do not assume normality but the straightforward normal
assumption appears to work fairly well to diagnose problems for all
cases presented here.  It is important to note that the number of
standard deviations a number is from the expected result is not
necessarily a measure of the size of the error.  Instead, it is a
measure of how certain we are of the error as we may be measuring
either a very small error with extremely high numerical precision or a
large error with little precision.

\subsection{Choosing the parameter gap}
We note that the relationship in Eq.~\ref{eq:canonicaltest} is true
for any choice of the temperatures $\beta_1$ and $\beta_2$.  However,
if $\beta_1$ and $\beta_2$ are very far apart then the two probability
distributions $P(E|\beta_1)$ and $P(E|\beta_2)$ will not be well determined over
any range of $E$ in any simulation of reasonable length.  If, on the
other hand $\beta_1 = \beta_2$, no information can be obtained
because the simulations will be statistically identical. If the two
simulations are not statistically identical there are deeper problems
to worry about than if the simulations are ensemble consistent!

Coming in from these two limits, if $\beta_1$ and $\beta_2$ are
moderately far apart, small-sample noise from the extremes of the
distribution will make it difficult to determine the deviations from
$\beta_2 - \beta_1$.  If $\beta_1$ and $\beta_2$ are too close
together, even the relatively small statistical noise at the centers
of the distributions will swamp out the information contained in the
very slight difference between the user-specified temperature gap and
the simulation's actual value for $\beta_2 - \beta_1$.  There should
therefore be some ideal range of temperature gaps giving the most
statistically clear information about deviations from ensemble
consistency.  We will examine specific choices of this gap for
different systems in this study.

\subsection{Sampling from the canonical ensemble with a harmonic oscillator}

To study these ensemble validity tests in practice, we first examine
a toy model, sampling from a $D$-dimensional harmonic oscillator. We
then use this model to demonstrate the use of this method to identify
simulation errors.

For a $D$-dimensional harmonic oscillator with an equal spring
constant $K$ in each dimension and equilibrium location $x_{i,0}$ in
each direction, the total potential energy of the system is $E =
\frac{1}{2} K\sum_{i=1}^{D}(x_i-x_{i,0})^2$.  The partition function
for this model is $Q(\beta) = (\frac{2\pi}{\beta K})^{D/2}$, meaning
the free energy is $A(\beta) = - (D/2\beta) \ln[ (2 \pi) / (\beta
  K)]$, and the probability of a given configuration $\vec{x}$ is
\[
P(\vec{x}|\beta) = \left(\frac{\beta K}{2 \pi}\right)^{D/2} \exp\left(-\frac{\beta K}{2} \sum_i^{D}|x_i-x_{i,0}|^2\right).
\] 
 
For this exercise, we set $x_{i,0}=0$ for all $i$ for simplicity, and
choose $D=20$. We specifically do not choose $D=1$, because it can
give results that may not be typical for other choices of dimensions.
For $D=1$, the density of states $\sigma(E)$ is constant for this
choice of $E$, i.e. $P(E) \propto \exp(-\beta E)$ for all spring
constants. Unlike most physical densities of states, in this case
$E=0$ has nonzero probability for all temperatures, which means
samples from all temperatures have nonnegligible overlap. Harmonic
oscillators with $D \gg 1$ have $\Omega(E)=0$ at $E=0$ and then
rapidly increasing as $E$ increases, much more characteristic of more
typical physical systems.

\subsubsection{Testing for ensemble validity with a toy system with simulation noise}

Using this toy system, we collect samples with $K=1$ and $\beta = 1.3$
and $0.7$ (the specific choice of temperature gap is explained later).
After generating the samples, we add random noise $\delta E = \nu
\left| N(0,1) \right|$, where $N(0,1)$ is a Gaussian random variate
with mean zero and standard deviation 1, and $\nu$ is some small
positive constant.  The addition of random noise allows us to test the
ability of the algorithm to identify simple errors in the energy
distributions.  In each case, we carry out 200 independent repetitions
of the procedure, each time with 500,000 samples from each of the two
different temperature distributions.  We note that this particular
type of error means that the data are generated with the correct
probability, but their energies are stored incorrectly.  This pattern
might not be typical of actual errors observed in molecular
simulations, but serves as a useful starting point for characterizing
the sensitivity of this procedure.  The results are shown as a
function of noise in Table~\ref{table:harmonic_noise}, with 0.6 the
exact result for $\beta_2 - \beta_1$.  The average energy of each
distribution is $-\frac{\partial \ln Q}{\partial \beta}=
\frac{D}{2\beta}$ or 16.667 in this specific case.  We examine the
linear, nonlinear, and maximum likelihood fits, with the error
calculated by the analytical estimates, sample standard deviations
over 200 repetitions, and bootstrap sampling using 200 bootstrap
samples from the first of the 200 repetitions.

\begin{table}[tbp]
\resizebox{16.5cm}{!}{
\begin{tabular}{|ccccccc|}
\hline
             &        $\nu$         &   0.0                         & 0.005                     & 0.0075                    & 0.01                      &                           0.02  \\
\hline
                & linear          &     0.6006 $\pm$ 0.0012 (0.5) & 0.5955 $\pm$ 0.0012 (3.7) & 0.5927 $\pm$ 0.0012 (6.0) & 0.5924 $\pm$ 0.0012 (6.3) &  0.5841 $\pm$    0.0012 (13.3)  \\ 
analytic        & nonlinear       &     0.6028 $\pm$ 0.0013 (2.3) & 0.6019 $\pm$ 0.0012 (1.6) & 0.5969 $\pm$ 0.0012 (2.5) & 0.5916 $\pm$ 0.0012 (6.9) &  0.5899 $\pm$    0.0012 (8.3)   \\
(single sample) & max. likelihood &     0.6016 $\pm$ 0.0012 (1.3) & 0.5973 $\pm$ 0.0012 (2.3) & 0.5939 $\pm$ 0.0012 (5.2) & 0.5936 $\pm$ 0.0012 (5.5) &  0.5850 $\pm$    0.0012 (13.0)  \\
\hline             
                & linear          &     0.5991 $\pm$ 0.0012 (0.8) & 0.5958 $\pm$ 0.0012 (3.6) & 0.5941 $\pm$ 0.0012 (4.8) & 0.5922 $\pm$ 0.0012 (6.5) &  0.5838 $\pm$    0.0011 (14.9) \\
200 replicates  & nonlinear       &     0.5991 $\pm$ 0.0031 (0.3) & 0.5964 $\pm$ 0.0030 (1.2) & 0.5946 $\pm$ 0.0031 (1.8) & 0.5931 $\pm$ 0.0031 (2.3) &  0.5870 $\pm$    0.0029 (4.4) \\
                & max. likelihood &     0.6001 $\pm$ 0.0012 (0.1) & 0.5968 $\pm$ 0.0011 (2.9) & 0.5950 $\pm$ 0.0012 (4.2) & 0.5932 $\pm$ 0.0012 (6.0) &  0.5848 $\pm$    0.0011 (14.2) \\
\hline
                & linear          &     0.5999 $\pm$ 0.0013 (0.0) & 0.5953 $\pm$ 0.0012 (4.1) & 0.5927 $\pm$ 0.0012 (6.2) & 0.5922 $\pm$ 0.0013 (6.0) &  0.5839 $\pm$    0.0013 (12.7) \\
200 bootstraps  & nonlinear       &     0.6019 $\pm$ 0.0034 (0.6) & 0.6017 $\pm$ 0.0031 (0.6) & 0.5940 $\pm$ 0.0029 (1.0) & 0.5915 $\pm$ 0.0032 (2.7) &  0.5890 $\pm$    0.0031 (3.1) \\
                & max. likelihood &     0.6013 $\pm$ 0.0012 (1.1) & 0.5972 $\pm$ 0.0011 (2.5) & 0.5950 $\pm$ 0.0012 (5.3) & 0.5935 $\pm$ 0.0013 (5.2) &  0.5848 $\pm$    0.0012 (12.4) \\
\hline
\end{tabular}
}
\caption{{\bf All fitting forms are sensitive determinants of noise in
    the energy.} Number of standard deviations from the true slope for
  the given error estimate method are shown in parentheses.  All
  fitting forms (linear, nonlinear, maximum likelihood) in combination
  with all estimators of the error (analytic, independent replicas,
  and bootstrap sampling) are sensitive determinants of noise amount
  ($nu$) in the energy. Nonlinear fitting is somewhat less useful, as
  the nonlinear analytical error estimate does not match the sample
  and bootstrap error estimates, in contrast to linear and maximum
  likelihood analytical error estimates.  The true $\beta_2 - \beta_1=
  0.6$.
\label{table:harmonic_noise}}
\end{table}

In all cases in Table~\ref{table:harmonic_noise}, bootstrap sampling
closely matches the standard sample error from 200 independent
samples, suggesting that bootstrap error estimation is likely to be as
effective as independent sampling to identify ensemble errors, as was
also observed in previous free energy
calculations.~\cite{Shirts2011}. Additionally, the analytical error
estimates for linear and maximum likelihood fitting closely match the
sample standard deviation.  This is particularly encouraging because it
means that single pairs of simulations are enough to calculate error
estimates robustly.

\begin{figure}[tbp]
\noindent
\begin{tabular}{cc}
\includegraphics[width=0.5\columnwidth]{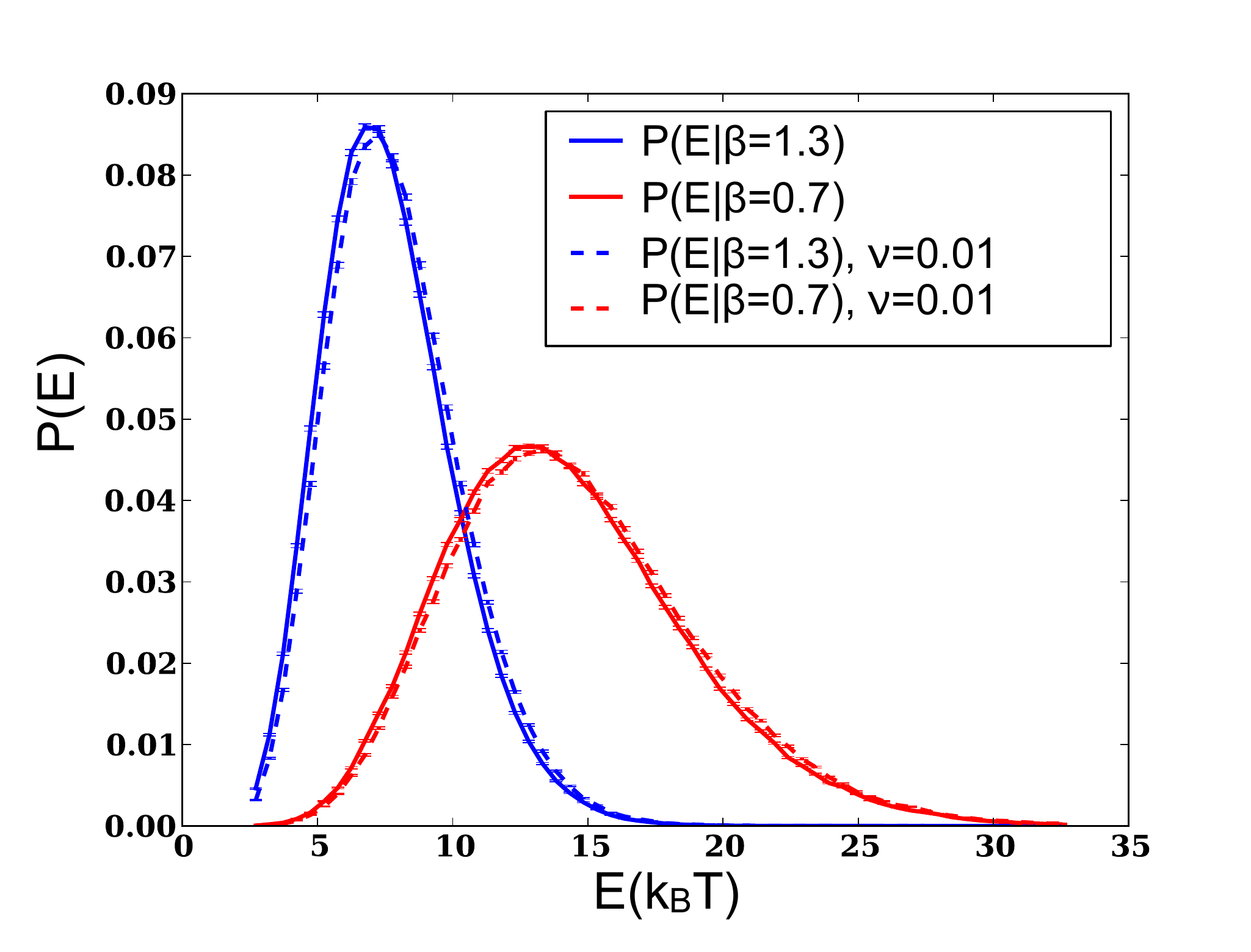} & \includegraphics[width=0.5\columnwidth]{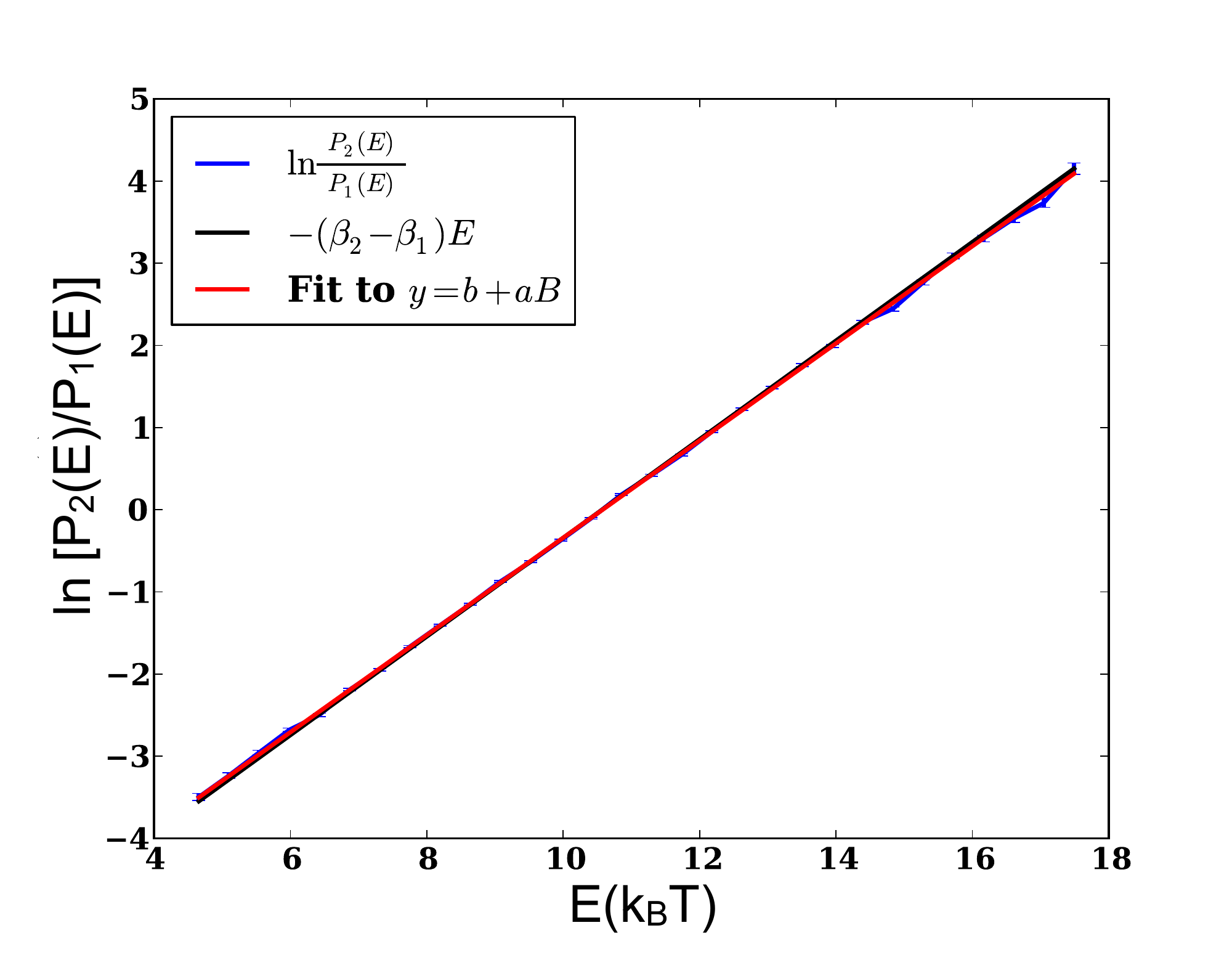} \\
a  &  b 
\end{tabular}
\caption{{\bf Model energy distributions and discrimination of error.}
  A small amount of noise (0.06\% of the average energy) is added to
  each sample.  Such differences affect the distribution minimally
  (a), and are difficult to tell from random noise in the (b) linear
  fits, but fitting quantitatively to the distribution reveals that
  the deviation in the distribution from analytical results is 5-7
  standard deviations beyond that which would be expected. The system
  is the same as used for Table~\ref{table:harmonic_noise} with
  error scale $\nu=0.01$.
\label{figure:discriminate_error}}
\end{figure}

Nonlinear fitting is somewhat less useful, as the nonlinear analytical
error estimates appear to noticeably underestimate the actual error,
as determined by the sample standard deviation over 200 repetitions.
The statistical error in nonlinear fitting is larger than the error in
the linear and maximum likelihood estimates, possibly because of a
magnified effect of small sample errors. However, all fitting forms
(linear, nonlinear, maximum likelihood) in combination with all
estimators of the error (analytic, independent replicas, and bootstrap
sampling) are relatively sensitive determinants of noise in the
energy. Deviations of more than 3$\sigma$ occur consistently for $\nu$
as low as $0.0075$, or less than 1\% of $k_{B}T$, demonstrating that
these errors have become statistically significant.  Even with
$\nu=0.01$, where the slope is between 5 and 7 standard deviations
from the true slope, the visual difference between estimates becomes
virtually unnoticeable, both for the actual distributions and the
ensemble validation fit, as seen in
Fig.~\ref{figure:discriminate_error}.  The ability to sensitively
identify errors that cannot be directly visualized demonstrates the
utility of this quantitative approach.  Overall, it appears that
maximum likelihood error estimates are the best method to use, as
discretization errors due to poor histogram choice will not
matter. However, linear fitting also appears robust, at least for this
system.

The ensemble validation relationship is true for all choices of
$\beta_1$ and $\beta_2$ but as discussed, for finite numbers of
samples, there are problems with choices of $\beta_1 - \beta_2$ that
are either too large or two small.  For large slope, small sample
error in the tails dominates; for small slopes, the small magnitude of
the slope becomes difficult to distinguish even at the moderate levels
of statistical error occurring near the peaks of the energy
distributions.  In Table~\ref{table:harmonic_noise}, we use a fixed
difference in temperatures. Can we identify an optimal temperature
difference to detect error?  For this exercise, we select a fixed low
level of random error ($\nu = 0.01$), and vary the slope
$\beta_{2}-\beta_{1}$ with the average $\frac{1}{2}(\beta_{1} +
\beta_{2})$ fixed at 1, using the analytical estimate from the maximum
likelihood parameter estimation, and again using 500,000 samples from
each distribution.

\begin{table}[tbp]
\resizebox{8.5cm}{!}{
\begin{tabular}{|ccccc|}
\hline
$\beta_{2}$ & $\beta_{1}$ & $\beta_2-\beta_1$  & Estimated  $\beta_2-\beta_1$ & $\sigma$ deviation \\
  1.05 &   0.95 &  0.1 &   0.0993 $\pm$    0.0006 & 1.1 \\ 
  1.10 &   0.90 &  0.2 &   0.1981 $\pm$    0.0007 & 2.7 \\ 
  1.15 &   0.85 &  0.3 &   0.2970 $\pm$    0.0008 & 3.9 \\ 
  1.20 &   0.80 &  0.4 &   0.3960 $\pm$    0.0009 & 4.7 \\ 
  1.25 &   0.75 &  0.5 &   0.4948 $\pm$    0.0010 & 5.2 \\ 
  1.30 &   0.70 &  0.6 &   0.5936 $\pm$    0.0012 & 5.4 \\ 
  1.40 &   0.60 &  0.8 &   0.7913 $\pm$    0.0017 & 5.1 \\ 
  1.50 &   0.50 &  1.0 &   0.9907 $\pm$    0.0027 & 3.5 \\ 
  1.60 &   0.40 &  1.2 &   1.1930 $\pm$    0.0047 & 1.5 \\ 
  1.70 &   0.30 &  1.4 &   1.3916 $\pm$    0.0100 & 0.8 \\ 
\hline
\end{tabular}
}
\caption{{\bf Optimizing temperature spacing to improve error
    detection in the distribution.} Deviation from the correct slope
  of the log ratio of the energy distributions as a function of
  increasing distance between the two distributions, as measured by
  the magnitude of $\beta_2 - \beta_1$, with fixed noise. The ability
  to discriminate the error in $\beta_2 - \beta_1$ reaches an optimum
  at intermediate separation of
  distributions.\label{table:noise-temperature-gap} }
\end{table}
For fixed noise in the energy function, we see in
Table~\ref{table:noise-temperature-gap} the number of standard
deviations from the true slope to the observed slope as a function of
energy gap. The ability to discriminate the error in $\beta_2 -
\beta_1$ is lower for both very large and very small temperature gaps,
though there is a relatively broad range near the middle where the
sensitivity of the test, measured in the number of standard deviations
the measured slope is from the true slope, is relatively constant.

Examining the energy distributions at maximum error discrimination
point ($\beta_1 = 0.6, \beta_2 =1.3$) we find that the difference
between the centers of the distributions ($14.3 k_B T - 7.7 k_B T =
6.6 k_B T$ ) is approximately equal to the sum of the standard
deviations of the distributions ($4.5 k_B T + 2.4 k_B T = 6.9 k_B
T$). This suggest a general rule-of-thumb that we can maximize the
ability to identify errors by using temperatures separated by
approximately the sum of the standard deviations of the distributions.
The precise value of the difference will not matter particularly in
most cases as long as we are somewhat near the optimum.  With less
data, we might err on the side of using a slightly smaller gap to
guarantee good overlap in the distributions near $E=0$.

This rule is simply intended as a guideline, as some sources of error
might show up preferentially in the tails and thus require larger
temperature gaps to observe, but provides a useful starting criteria.
One example of a physical system which violates this rule is a 1-D
harmonic oscillator, which has a constant density of states
$\Omega(E)$. Although the statistical error does indeed increase with
decreasing overlap, the slope increases faster, and thus sensitivity
to statistical error always increases with increasing temperature gap.
With fixed error, the sensitivity with noise magnitude $\nu=0.02$
increases from less than one standard deviation for
$\beta_2-\beta_1=0.1$ to over five standard deviations for
$\beta_2-\beta_1 = 1.8$.  However, this case is atypical, because the
density of states is a constant with the maximum probability always at
$E=0$, so that even when the temperatures are very different there is
still nonnegligible overlap in the distributions.

To apply this rule, we still need to estimate the standard
distributions of the two distributions. If we assume the variance in
energy (and therefore the value of the heat capacity) does not change
very much over the relative narrow range of temperature spacings used
to perform ensemble validation, then the distributions will also be
the same. We can estimate the width $\sigma$ of the distribution given
a known heat capacity $C_V$.  Specifically, $\sigma_E = T \sqrt{C_V
  k_B}$, so that for a temperature gap to result in a difference in
the centers of the energy distribution of $2\sigma_E$, we must have
$2\sigma_E = \frac{\partial E}{\partial T}\Delta T = C_V\Delta T$,
which reduces to $\Delta T/T = \sqrt{2k_B/C_V}$.  Alternatively, in
many cases it may make the most sense to run a short simulation at the
``center'' temperature $\frac{1}{2}\left(T_1 + T_2\right)$ to estimate
the variance, and we can use the equivalent relationship $\Delta T/T =
\frac{2k_BT}{\sigma_E}$ to identify a reasonable temperature gap
$\beta_2 - \beta_1$ for simulations of a specific system.

\subsection{Isobaric-isothermal ensembles}
Our discussion up to this point has been restricted to NVT systems.
However, the same principles can also be applied to check the validity
of simulations run at constant temperature and pressure, and of
simulations run at constant temperature and chemical potential.  We
will analyze isobaric-isothermal simulations extensively in this
section.  We will not examine grand canonical simulations in this
paper, though we do include the derivations in
Appendix~\ref{appendix:grand}.

There are at least three useful ways we can analyze NPT simulations
for validity.  First, let us assume that we have two simulations run
at the same pressure but different temperatures.  Then the microstate
probabilities are:
\begin{eqnarray}
P(\vec{x},V|\beta,P) = \Delta(\beta,P)^{-1}\exp(-\beta E(\vec{x}) -\beta P V)
\end{eqnarray}
Where $\Delta(\beta,P)$ is the isothermal-isobaric partition function.
We then integrate out configurations with fixed instantaneous enthalpy
$H = E(\vec{x}) + PV$, where $\vec{x}$ here is shorthand for both
position and momentum variables, not the entire microstate specification.  We then have
\begin{eqnarray*}
P(H|\beta,P) &=& \frac{\beta P}{h^{3N}}\int_V \int_{\vec{x}} \delta[E(\vec{x})+PV-H]\Delta(\beta,P)^{-1}\exp(-\beta (E(\vec{x}) + P V) d\vec{x} dV \\
     &=& \frac{\beta P}{h^{3N}}\Omega^{\prime}(H,P)\Delta(\beta,P)^{-1}\exp(-\beta H)
\end{eqnarray*}
where $\Omega^{\prime}(H,P)$ is a density of states counting the
number of states with a given value of $H = E+PV$, and is explicitly a
function of $P$, but not $\beta$. The prefactor of $\beta P$ comes
from the requirement to cancel the units in the integral, ignoring
factors of $N$ relating to distinguishability of particles, which will
cancel in the ratio of distributions in all cases. Because both simulations have the same pressure,
we arrive directly at a new ensemble validation relationship:
\begin{eqnarray}
\frac{P(H|\beta_2,P)}{P(H|\beta_1,P)} &=& \frac{\beta_1\Delta(\beta_1,P)}{\beta_2 \Delta(\beta_2,P)}\exp(-[\beta_2 - \beta_1] H) \\
\ln \left(\frac{P(H|\beta_2,P)}{P(H|\beta_1,P)}\right) &=& \ln (\beta_1/\beta_2) + \left[\beta_2 G_2 - \beta_1 G_1\right] - [\beta_2 - \beta_1] H).
\label{eq:enthalpytest}
\end{eqnarray}
The exact same ensemble validation statistical tests can be
applied with $H$ in place of $E$ and the Gibbs free energy (or free
enthalpy) $G$ plus a small correction factor in the place of $A$.

We can also look at the probability of the volume alone by
integrating out the energy $E$ at fixed volume:
\begin{eqnarray}
P(V|\beta,P_1) &=& \frac{\beta_1 P_1}{h^{3N}} \Delta(\beta,P_1)^{-1}Q(\beta,V)\exp(-\beta P_1 V)  \nonumber \\
P(V|\beta,P_2) &=& \frac{\beta_2 P_2}{h^{3N}} \Delta(\beta,P_2)^{-1}Q(\beta,V)\exp(-\beta P_2 V)  \nonumber \\
\frac{P(V|\beta,P_2)}{P(V|\beta,P_1)} &=& \frac{P_1\Delta(\beta,P_1)}{P_2\Delta(\beta,P_2)}\exp(-\beta [P_2 - P_1] V) \\
\ln \frac{P(V|\beta,P_2)}{P(V|\beta,P_1)} &=& \ln (P_1/P_2) +\left[\beta (G_2 - G_1)\right]- \left[\beta(P_2 - P_1) V \right]
\label{eq:volumetest}
\end{eqnarray}
We can then use the same techniques already described with
$\Delta(\beta,P_1)$ in the place of $Q(\beta,V)$, $P_1$ and $P_2$ in
the place of $\beta_1$ and $\beta_2$ and $\beta V$ in the place of
$E$.

Finally, we can treat the joint probability distributions with both $V$ and $E$ varying independently:
\begin{eqnarray}
P(V,E|\beta_1,P_1) &=& \frac{\beta_1 P_1}{h^{3N}}\Omega(V,E)\Delta(\beta_1,P_1)^{-1}\exp(-\beta_1 E- \beta_1 P_1 V) \nonumber \\
P(V,E|\beta_2,P_2) &=& \frac{\beta_2 P_2}{h^{3N}}\Omega(V,E)\Delta(\beta_2,P_2)^{-1}\exp(-\beta_2 E- \beta_2 P_2 V) \nonumber \\
\frac{P(V,E|\beta_2,P_2)}{P(V,E|\beta_1,P_1)} &=& \frac{\beta_1 P_1\Delta(\beta_1,P_1)}{\beta_2 P_2\Delta(\beta_2,P_2)}\exp([\beta_2-\beta_1]E + [\beta_2 P_2 - \beta_1 P_1] V) \\
\ln \frac{P(V,E|\beta_2,P_2)}{P(V,E|\beta_1,P_1)} &=& \ln (\beta_1 P_1/\beta_2 P_2) + \left[\beta_2 G_2 - \beta_1 G_1\right] - \left[\beta_2-\beta_1\right]E - \left[\beta_2 P_2 - \beta_1 P_1\right] V
\label{eq:jointtest}
\end{eqnarray}
We can apply most of the same methods described previously with slight
modifications for the additional dimensions.  For example, when
fitting log ratio of the distributions, we must now perform a
multilinear fit in $V$ and $E$.  Multiple variable nonlinear fitting
can also be employed. However, in both cases, we can quickly run into
numerical problems because of the difficulty of populating
multidimensional histograms with a limited number of samples, making
discretization error worse.  The maximum likelihood method, which
already appears to be the most reliable method for estimating single
variables, does not require any histograms and thus is free from
discretization error in any dimension. In examining joint variation in
$E$ and $V$ in this study we therefore focus on only the maximum
likelihood method.
 
For maximum likelihood maximization we again need to clarify what the
free variables are in order to fix the form of the probability distribution.
The first two are $\Delta G = G_2 - G_1$, setting $G_1 + G_2 = 0$, and $\Delta
\beta = \beta_2 - \beta_1$, setting $(\beta_1 + \beta_2)/2 =
\beta_{\mathrm{ave}} = \mathrm{const}$, as before.  By analogy, we set
$(P_1 + P_2)/2 = P_{\mathrm{ave}}$, with the variable
$\Delta P = P_1 - P_2$. Both $\beta_{\mathrm{ave}}$ and
$P_{\mathrm{ave}}$ are then set at the averages of the applied $\beta$
and $P$ of the two simulations. We then find that:
\begin{eqnarray}
(\beta_2 P_2 - \beta_1 P_1) &=& \frac{1}{2}\left((\Delta \beta )(P_2+P_1) + (\beta_2+\beta_1)(\Delta P )\right) \nonumber \\
                            &=& (\Delta \beta)P_{\mathrm{ave}} + \beta_{\mathrm{ave}}(\Delta P) 
\end{eqnarray}
The explicit maximum likelihood equations for enthalpy, volume, and joint energy and volume are then:
\begin{eqnarray}
\ln \frac{P(H|\beta_2,P)}{P(H|\beta_1,P)} &=& \beta_{\text{ave}}(\Delta G) -(\Delta \beta)H \\
\ln \frac{P(V|\beta,P_2)}{P(V|\beta,P_1)} &=& \beta(\Delta G -(\Delta P) V) \\
\ln \frac{P(E,V|\beta_2,P_2)}{P(E,V|\beta_1,P_1)} &=& \beta_{\text{ave}}(\Delta G) - (\Delta \beta) E -  (\Delta \beta)P_{\mathrm{ave}}V - (\Delta P)\beta_{\mathrm{ave}}V  \nonumber \\
                              &=& \beta_{\text{ave}}(\Delta G) - (\Delta \beta) (E + P_{\mathrm{ave}}V) - \beta_{\mathrm{ave}}(\Delta P)V 
\end{eqnarray}
omitting the unchanged prefactors involving logarithms of the ratios
of the known intensive variables $\beta_1$, $\beta_2$, $P_1$ and
$P_2$.  In general, we can ignore this term because we usually don't
care about the exact value of the free energy difference $\Delta G$
between the paired simulations, and so therefore do not need to break
the constant term down into its components.

\subsection{Sampling from the isobaric-isothermal ensemble for a toy problem}

To better understand how to validate the volume ensemble, we examine a
toy model sampling from a modified harmonic oscillator potential.  In
this case, the harmonic spring constant is increased by decreasing
system volume in order to add a $PV$ work term to the system. We set the
harmonic force constant $K = (a/V)^2$, and for simplicity set $x_0=0$.
This means that $\Delta(P,\beta) = \int_V Q(V,\beta) \exp(-\beta V)
dV$ which gives
\begin{eqnarray*}
\Delta(P,\beta) &=& (\beta P)^{-2}\sqrt{\frac{2\pi}{a^2\beta}} \\
P(x,V|\beta,P) &=& a(\beta P)^2\sqrt{\frac{\beta}{2\pi}} \exp\left(-\frac{\beta a^2x^2}{2V^2} - \beta P V\right)
\end{eqnarray*}

We use the Gibbs sampler~\cite{GemanS1984} to generate configurations
from the joint distribution $P(E,V)$ by alternating sampling in
$P(E|V)$ and $P(V|E)$. To sample randomly from $P(x|V)$, we observe
that $x$ will always be distributed as a Gaussian, with standard
deviation $\sigma=\sqrt{K/\beta} = (V/a)\beta^{-1/2}$.  To perform
conditional sampling in the system volume dimension, we must sample
according to the conditional distribution $P(V|x_i) \propto
\exp\left(-\beta \frac{a^2 x_i^2}{2V^2} -\beta PV\right)$.  This is
not a typical continuous probability family so there is no simple
formula for generating samples from this distribution. However, we
note that the distribution is strictly less than $M\exp(-\beta P V)$,
where $M$ is the ratio of the normalizing constant for the exponential
distribution and the normalizing constant for the exponential plus the
harmonic term. We can then sample $V$ from the exponential
distribution $\exp(-\beta PV)$ and perform rejection sampling to
sample from the strictly smaller desired distribution $P(V|x)$.
Initially, it appears that the smaller the difference between the two
distributions (i.e. the smaller $\frac{\beta a^2x_i^2}{2V^2}$) is, the
more efficient the sampling will be.  However, because $x_i$ is
generated from a Gaussian distribution, $\langle x^2 \rangle =
\frac{\beta V^2}{a^2}$, then the average efficiency reduction factor
becomes $\exp(-\beta/2)$, independent of $P$ or $a$, so the acceptance
ratio is only significantly affected by the temperature.

\subsubsection{NPT Model System Results}
\begin{figure}[tbp]
\noindent
\begin{tabular}{cc}
\includegraphics[width=0.5\columnwidth]{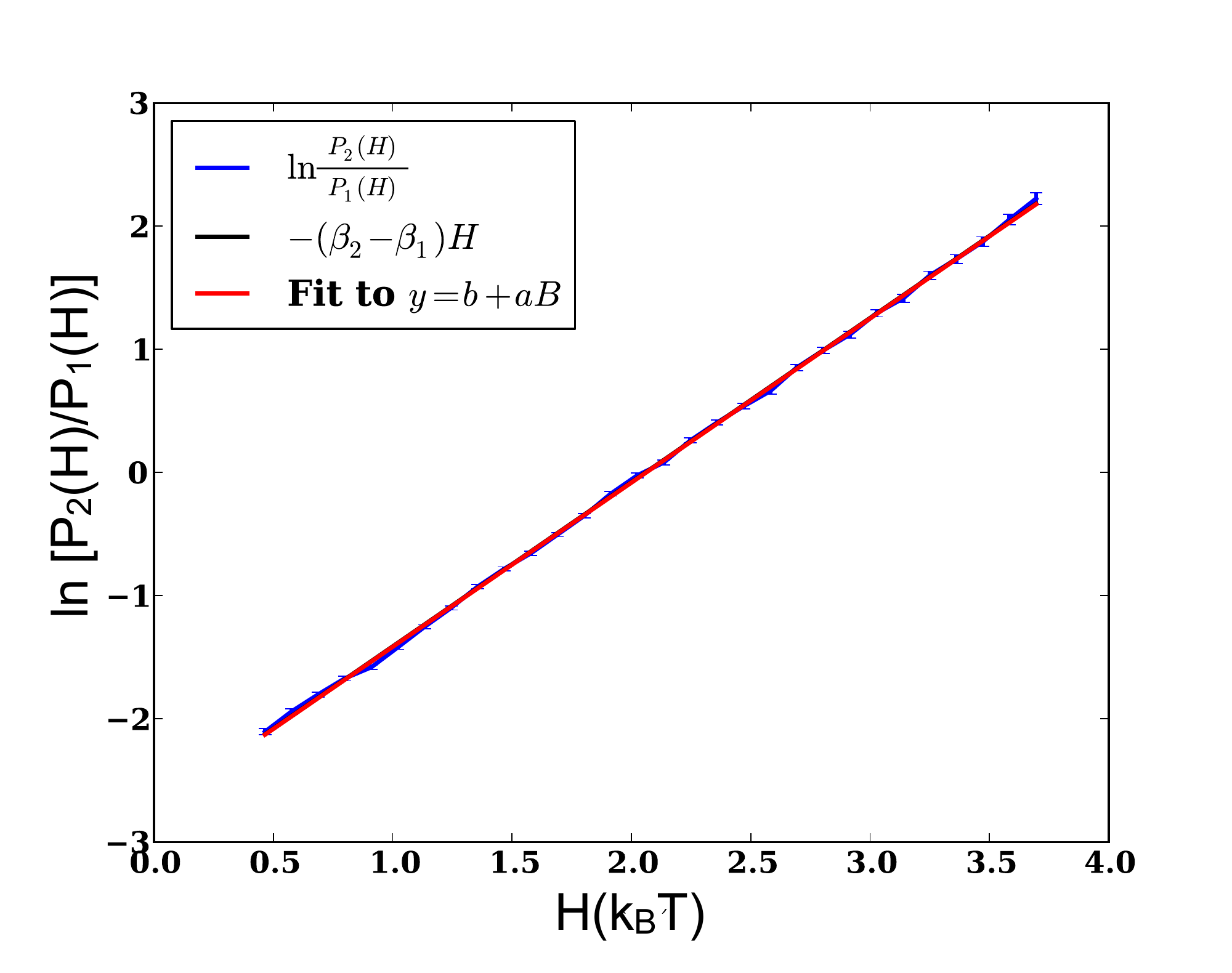} & \includegraphics[width=0.5\columnwidth]{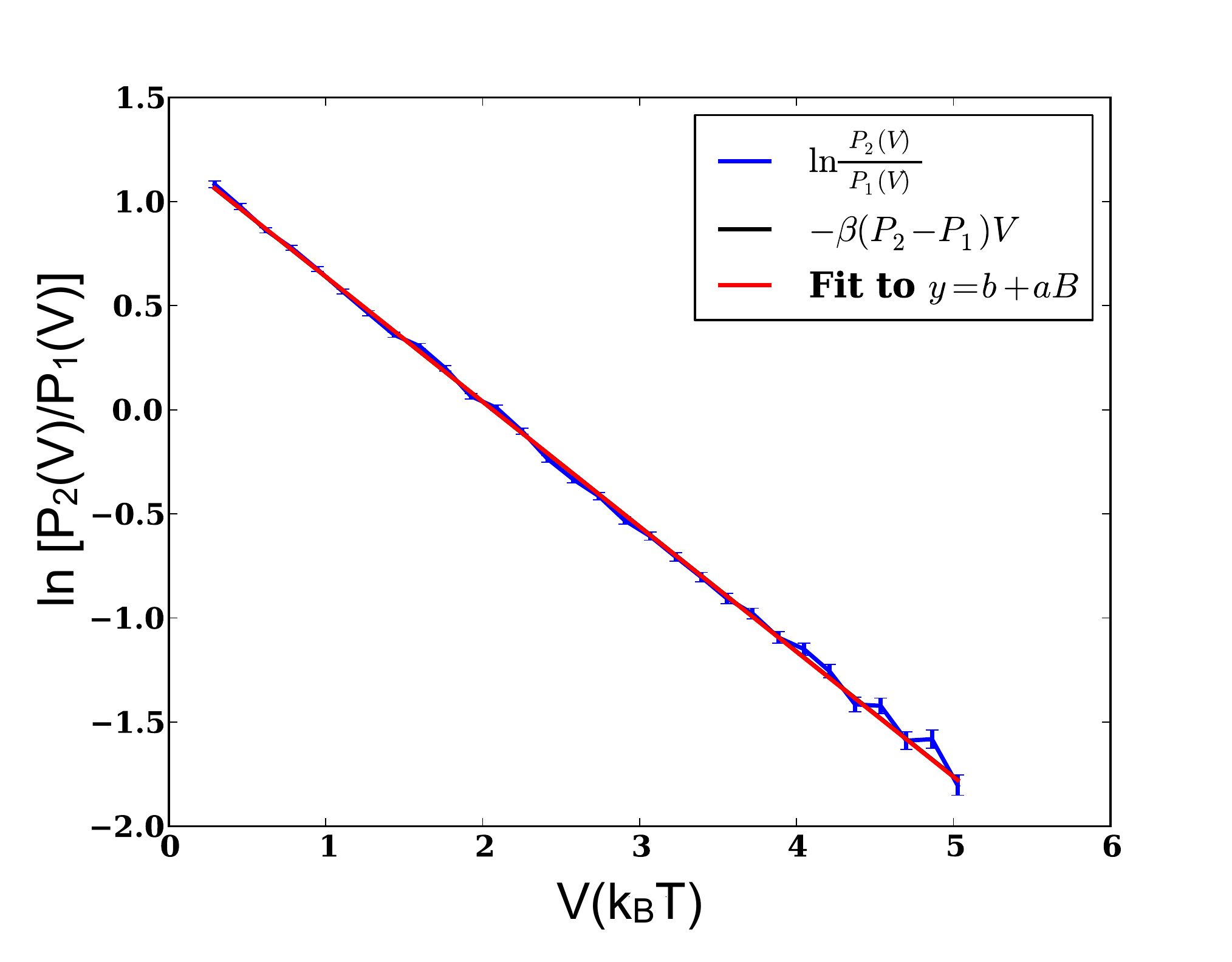} \\
a  &  b 
\end{tabular}
\caption{{\bf Validation of distributions for harmonic oscillators
    with pressure} We can accurately validate the isothermal-isobaric
  distributions of enthalpy (a) and volume (b) for our harmonic
  oscillator toy problem with a volume dependent spring constant.\label{fig:harmonic_pressure}}
\end{figure}

For all tests, we generate 250,000 samples
from each of the paired distributions.  To examine the enthalpy, we
pick $\beta_1 = 2/3$, $\beta_2 = 2$, and $P_1=P_2=1$, and using the
maximum likelihood method we estimate $\beta_2 -\beta_1 = 1.3341 \pm
0.0040$, only 0.2 quantiles from the true answer of 4/3 (see
Fig.~\ref{fig:harmonic_pressure}a for the linear plot).  To validate
the volume sampling, we pick $\beta_1 = \beta_2 = 1.0$ and $P_1=1.3$
and $P_2=0.7$, and find that $\beta(P_2-P_1) = -0.6013 \pm 0.0025$,
0.53 quantiles from the true answer of -0.6 (see
Fig.~\ref{fig:harmonic_pressure}b) for the linear plot.
Finally, when we examine the joint variation of energy and volume, we
use $\beta_1=0.6$, $\beta_2 = 0.8$, $P_1 = 0.8$ and $P_2 = 1.2$, which
gives us $0.20035\pm0.00318$ for the slope $(\beta_2-\beta_1)$ and
$-0.48129 \pm 0.00185$ for the slope $\beta_2 P_2 - \beta_1 P_1$,
which are 0.1 and 0.7 quantiles from the true answers $0.2$ and
$-0.48$, respectively.  We see that indeed these equations properly
capture entropy and volume distributions.

\subsubsection{Picking intervals for enthalpy and volume tests}

In the NPT case with differing temperatures and constant pressure, the
instantaneous enthalpy $E + PV$ takes the place of the energy and a
two standard deviation temperature gap will mean choosing temperatures
separated by $\sqrt{2k_B/C_P}$, instead of $\sqrt{2k_B/C_V}$.  In the
case of an NPT simulation performed with constant temperature and at
differing pressures, we want $2\sigma_V = \Delta P\left(\frac{\partial
  V}{\partial P}\right)_T$. We can use the distribution of volume
fluctuations to find that $\left(\frac{\partial V}{\partial
  P}\right)_T = -\frac{\sigma_V^2}{k_BT}$. We therefore must have that
$|\Delta P| = 2k_BT/\sigma_V$, or in terms of the physical measurable
isothermal compressibility $\kappa_T =
-\frac{1}{V}\left(\frac{\partial V}{\partial P}\right)_T$, $\Delta P =
\sqrt{2\frac{k_B T}{V \kappa_T}}$.  Again, this is a guideline, not a
strict rule; short simulations at the simulation average can also be
useful to identify the spread of the distributions, as the answer must
only be in the right range.  For joint distributions, the analysis is
more complicated, but it seems reasonable to use $2\sigma$ in both
directions, perhaps erring on the low side to ensure sufficient
samples.

\section{Molecular systems}

\subsection{Kinetic energy and potential energy independently obey the ensemble validation equation}
In most molecular systems (for example, ones without applied magnetic
fields), the potential energy of the system is independent of the
velocities and masses of the particles.  Thus, the potential and
kinetic energy are separable, and we can write:
\begin{eqnarray*}
 P(E_{\mathrm{pot}}+E_{\mathrm{kin}}|\beta) &=& Q_{\mathrm{kin}}(\beta)^{-1}Q_{\mathrm{pot}}(\beta)^{-1} \Omega(E_{\mathrm{pot}})\Omega(E_{\mathrm{kin}}) \exp(-\beta E_{\mathrm{pot}})\exp(-\beta E_{\mathrm{kin}}) \\
                          &=& \left[Q_{\mathrm{kin}}(\beta)^{-1}\Omega(E_{\mathrm{kin}})\exp(-\beta E_{\mathrm{kin}})\right]\left[ Q_{\mathrm{pot}}(\beta)^{-1} \Omega(E_{\mathrm{pot}}) \exp(-\beta E_{\mathrm{pot}})\right] \\
                          &=& P(E_{\mathrm{pot}}|\beta)P(E_{\mathrm{kin}}|\beta)
\end{eqnarray*}
The separability of the density of states occurs again because the momenta can be sampled independently of the coordinates.
The ensemble validation algorithm is therefore valid for the kinetic and potential energies independently as well, so that:  
\begin{eqnarray}
\frac{P(E_{\mathrm{kin}}|\beta_2)}{P(E_{\mathrm{kin}}|\beta_1)} &=& \frac{Q_{\mathrm{kin}}(\beta_2)}{Q_{\mathrm{kin}}(\beta_1)}\exp(-[\beta_2-\beta_1]E_{\mathrm{kin}})\\
\frac{P(E_{\mathrm{pot}}|\beta_2)}{P(E_{\mathrm{pot}}|\beta_1)} &=& \frac{Q_{\mathrm{pot}}(\beta_2)}{Q_{\mathrm{pot}}(\beta_1)}\exp(-[\beta_2-\beta_1]E_{\mathrm{pot}})
\end{eqnarray}
In the case of kinetic energy, $Q_{\mathrm{kin}}$ is simply
$\prod_{i=1}^N \int_{-\infty}^{\infty}\exp(-\beta p_i^2/m_i)dp_i = 
\prod_{i=1}^N (\frac{m_i}{\pi \beta})^{3/2}$, meaning the probability ratio is:
\begin{eqnarray}
\frac{P(E_{\mathrm{kin}}|\beta_2)}{P(E_{\mathrm{kin}}|\beta_1)} &=& \left(\frac{\beta_2}{\beta_1}\right)^{3N/2}\exp([\beta_1-\beta_2]E_{\mathrm{kin}})\\
                                                    &=& \left(\frac{\beta_{\mathrm{ave}}+\Delta\beta}{\beta_{\mathrm{ave}}-\Delta\beta}\right)^{3N/2}\exp(-\Delta\beta E_{\mathrm{kin}})
\end{eqnarray}
which is now in terms of the single free parameter $\Delta \beta =
\beta_2 - \beta_1$ rather than two parameters.  Note that this is true
for both identical and non-identical particles, since the mass terms
will cancel out for all $i$. In the case of kinetic energy, we can
obtain a distribution for each distribution alone, because the kinetic
energy is simply the sum of $3N$ random normal variables with standard
deviations $(m_i)^{-1/2}p_i$, and is thus a $\chi^2$ distribution with
$3N$ (minus any center of mass variables removed from the simulation)
degrees of freedom (DOF).  For more than 60 DOF, corresponding to
about 20 particles, the $\chi^2$ distribution is essentially
indistinguishable from a normal distribution with mean equal to the
sum of the means of the individual distributions, which in this case
is simply the average kinetic energy.  By equipartition the total kinetic
energy will simply be $\frac{3N}{2\beta}$. The standard deviation can
be computed by noting that the $\sigma^2 = k_B T^2 C_V$, and that the
heat capacity due to the kinetic energy is the ideal gas heat
capacity, $\frac{3Nk_B}{2}$.  Thus $\sigma^2 = \frac{3N}{2\beta^2}$,
and
\[
P(E_{\mathrm{kin}}) = \frac{\beta}{\sqrt{3N\pi}}\exp \left(-\frac{\left(\beta E_{\mathrm{kin}}-\frac{3N}{2}\right)^2}{3N}\right)
\]
to high accuracy for any number of molecules typical in molecular
simulations. In the above formulas, $3N$ should be replaced by the
correct number of DOF if constraints are implemented or if any center
of mass degrees of freedom are removed.  Standard methods for testing
the normality of distributions with known means and standard
deviations can be used, such as inspecting Q-Q plots or the
Anderson-Darling test.~\cite{Anderson1952} If the number of degrees of
freedom is not available, as may be the case when one is analyzing
data provided by someone else, then this can be estimated from the
average of the kinetic energy by equipartition as $\langle
E_{\mathrm{kin}}\rangle = \frac{k_B T}{2}(\# DOF)$.  If the kinetic
energy is not equal to this value, then the reported temperature will
not even be correct, which should be noticed from simpler outputs of
the simulation before running any other more sophisticated analysis
like the procedures described in this paper.

The kinetic energy distribution, in addition to following the ensemble
validation formula, can therefore be checked directly as well, though
this does not seem to be common practice in molecular simulation
validation.  The potential energy formula can be used to either
validate the potential energies separately, or can be used for Monte
Carlo simulations, where only potential energies are defined.  It is
also possible to perform this separation in terms of ideal gas and
canonical partition functions, but it does not change the results, as
the volume is constant.

To obtain separability of kinetic and potential energies in an NPT
ensemble, we start by writing the isobaric-isothermal partition
function in terms of kinetic and potential energy portions of the
canonical partition functions, and note that the kinetic energy part is
independent of the volume.
\begin{eqnarray*}
 \Delta(\beta,P) &=& \beta P \int_V Q_{\mathrm{kin}}(\beta) Q_{\mathrm{pot}}(\beta,V) \exp(-PV) dV \\
 \Delta(\beta,P) &=& Q_{\mathrm{kin}} \beta P \int_V Q_{\mathrm{pot}}(\beta,V) \exp(-PV) dV \\
 \Delta(\beta,P) &=& Q_{\mathrm{kin}}(\beta) \Delta_{\mathrm{pot}}(\beta,V) 
\end{eqnarray*}
Then in terms of probabilities, we have:
\begin{eqnarray*}
P(E,V|\beta,P) &=& Q_{\mathrm{kin}}(\beta)^{-1} \Delta_{\mathrm{pot}}(\beta,V)^{-1}\exp(-\beta E_{\mathrm{kin}}-\beta E_{\mathrm{pot}} -\beta P V) \\
P(E_{\mathrm{kin}}|\beta,P) &=& Q_{\mathrm{kin}}(\beta)^{-1}\exp(-\beta E_{\mathrm{kin}}) \\   
P(E_{\mathrm{pot}},V|\beta,P) &=& \Delta_{\mathrm{pot}}(\beta,V)\exp(-\beta E_{\mathrm{pot}}-\beta P V) 
\end{eqnarray*}

This separation again makes it possible to validate NPT Monte Carlo
simulations by removing the kinetic energy.

\subsection{Molecular dynamics of Lennard-Jones spheres}

We next illustrate of the utility of the ensemble validation formula
for molecular simulations.  For this study, we used a simulation of
300 Lennard-Jones particles using a beta version of the Gromacs 4.6
simulation code compiled in double precision.  We used the Rowley,
Nicholson and Parsonage argon parameters for Lennard-Jones spheres
($\sigma = 0.3405$ nm, $\epsilon = 119.8$ K, $k_{B} =
0.996072\;\mathrm{kJ/mol}$),~\cite{Rowley1976} and simulated at
$\rho = 0.85 \rho_c$, meaning the box is of length 3.5328256 nm and $T
= 0.85 T_c = 135.0226$.  Velocity Verlet integration was used, with
the exception of the Gromacs stochastic integration method, which is
only defined for the leapfrog Verlet algorithm. Linear center of mass
momentum was removed every step, and a long range homogeneous
dispersion correction was applied to the energy. Unless otherwise
specified, a Lennard-Jones switch between 0.8 and 0.9 nm was used,
with a neighborlist at 1.0 nm, a neighborlist update frequency of 5
step, and a time step of 8 fs.  Temperature coupling algorithms were
carried out with a coupling constant of $\tau_T = 1.0$~ps.  A total of
62.5 million MD steps were simulated for all simulations, equivalent
to 500 ns with a 8 fs time step, with the last 490 ns used for
analysis.  Unless otherwise specified, the low and high temperatures
are $T = 132.915$ and $T = 137.138$ respectively, chosen to be
approximately 0.7 times the estimated ideal $\sigma$ gap from the rule
of thumb, using $C_V \approx 8.5$ kJ K$^{-1}$mol$^{-1}$ from a
preliminary simulation of the system.

\subsection{Molecular example: validating temperature control algorithms}
Using this Lennard-Jones system, we first examine temperature control
algorithms implemented in Gromacs: Bussi-Parrinello,~\cite{Bussi2007}
with stochastic scaling of the target temperature, Andersen
temperature control,~\cite{Andersen1980} a variant of Andersen
temperature control with the velocity of all atoms randomized at some
regular interval $\tau_t$, Nos\'{e}-Hoover,~\cite{Hoover1985}
stochastic dynamics, and Berendsen velocity
scaling.~\cite{Berendsen1984} All of these temperature control
algorithms are proven in theory to give the correct canonical
distribution in the limit of long time scales~\cite{note_hoover} with
the exception of the Berendsen temperature algorithm, which is known
to give an incorrect, overly narrow kinetic energy
distribution.~\cite{Morishita_2000,Lemak_1994,Golo_2002} We examine
the deviations of the total, potential, and kinetic energies, using
analytic errors from the maximum likelihood fits.  In this analysis we
will often use the $\Delta P$ and $\Delta T$ (from the maximum
likelihood expressions) to describe the deviations from the true
distribution to make them more intuitive.  We can calculate $\Delta T$
from $\Delta \beta$ by assuming an average $\beta_{\mathrm{ave}} =
\frac{1}{2}(\beta_1 + \beta_2)$, and calculating $T_2 =
k_B^{-1}(\beta_{\mathrm{ave}} + \Delta \beta/2)^{-1}$ and $T_1 =
k_B^{-1}(\beta_{\mathrm{ave}} - \Delta \beta/2)^{-1}$.
In all molecular simulations, we also compute the correlation times
$\tau$ of the energy observables, using the {\tt timeseries} module of
the pymbar code distribution~\cite{Shirts2008} and subsample the data
with frequency $2\tau+1$ to obtain uncorrelated samples.  We find that
for the kinetic energies alone, the correlation times are actually
artificially short when using the algorithm in the {\tt timeseries}
module, which only integrates out to the first crossing of the
x-axis. We therefore in this study use the correlation times for the
potential energies, which are equal to longer than the correlation
times of the kinetic or total energies.  Subsampling more frequently
than required only affects the results by decreasing the statistical
accuracy due to collecting to few uncorrelated measurements, which for
a validation test is not as large a problem as significantly
undersampling the statistical error, which results in using correlated
data.  For the thermostat comparison, we use the subsampling
frequencies of 40 ps, which is the maximum among all methods, except
for the Andersen massive variant, for which we use 60 ps.

\begin{table}[tbp]
\resizebox{16.5cm}{!}{
\begin{tabular}{|l|cc|cc|cc|}
\hline
                      & \multicolumn{6}{c|}{True $\Delta T = 4.223$} \\ 
Thermostat            & \multicolumn{2}{c}{total} & \multicolumn{2}{c}{potential} & \multicolumn{2}{c|}{kinetic} \\
\hline
                      & Estimated $\Delta T$ & $\sigma$ deviation & Estimated $\Delta T$ & $\sigma$ deviation & Estimated $\Delta$ T & $\sigma$ deviation \\ 
\hline
         None (NVE)   &  \multicolumn{2}{c|}{N/A (constant)} &  4.388$\pm$0.115 &  1.4  &  3.048$\pm$0.112 & 10.5   \\
         Berendsen    &  9.369$\pm$0.122 &  42.2       &  4.606$\pm$0.086 &  4.5  & 29.034$\pm$0.364 & 68.3   \\
        Stochastic    &  4.172$\pm$0.066 &  0.8        &  4.098$\pm$0.081 &  1.6  &  4.251$\pm$0.091 &  0.3   \\
   Nos\'{e}-Hoover    &  4.197$\pm$0.067 &  0.4        &  4.220$\pm$0.082 &  0.03 &  4.186$\pm$0.090 &  0.4   \\
          Andersen    &  4.212$\pm$0.066 &  0.2        &  4.226$\pm$0.081 &  0.03 &  4.226$\pm$0.090 &  0.03  \\
Andersen (Massive)    &  4.188$\pm$0.079 &  0.4        &  4.176$\pm$0.097 &  0.5  &  4.217$\pm$0.107 &  0.06  \\
  Bussi-Parrinello    &  4.167$\pm$0.066 &  0.8        &  4.272$\pm$0.082 &  0.6  &  4.155$\pm$0.089 &  0.8   \\
\hline
\end{tabular}
}
\caption{{\bf Ensemble validation of different temperature control
    algorithms } All studied thermostats are consistent with a
  canonical ensemble, with the exception of the Berendsen thermostat,
  with deviations from the true slope generally 1 $\sigma$ or
  less. The true slope is 0.027865 $k_BT^{-1}$, equivalent to $\Delta
  T = 4.223$.  All errors are computed using the maximum likelihood
  method with the analytical error estimate.  NVE simulations also
  deviate from the canonical ensemble though the potential energy
  distributions do not statistically
  deviate.\label{table:thermostat_validation} }
\end{table}

\begin{figure}[tpb]
\noindent
\begin{tabular}{cc}
\includegraphics[width=0.5\columnwidth]{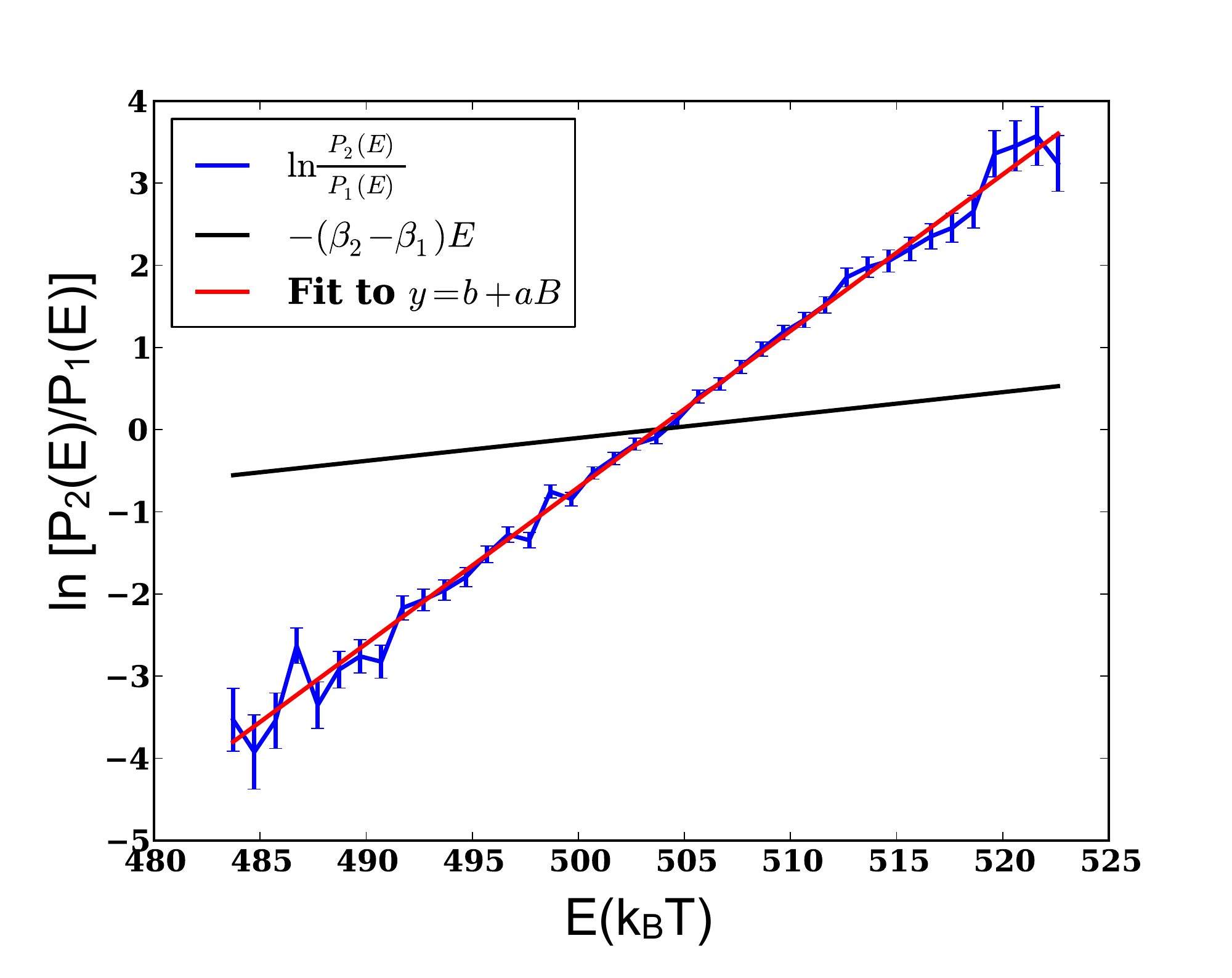} & \includegraphics[width=0.5\columnwidth]{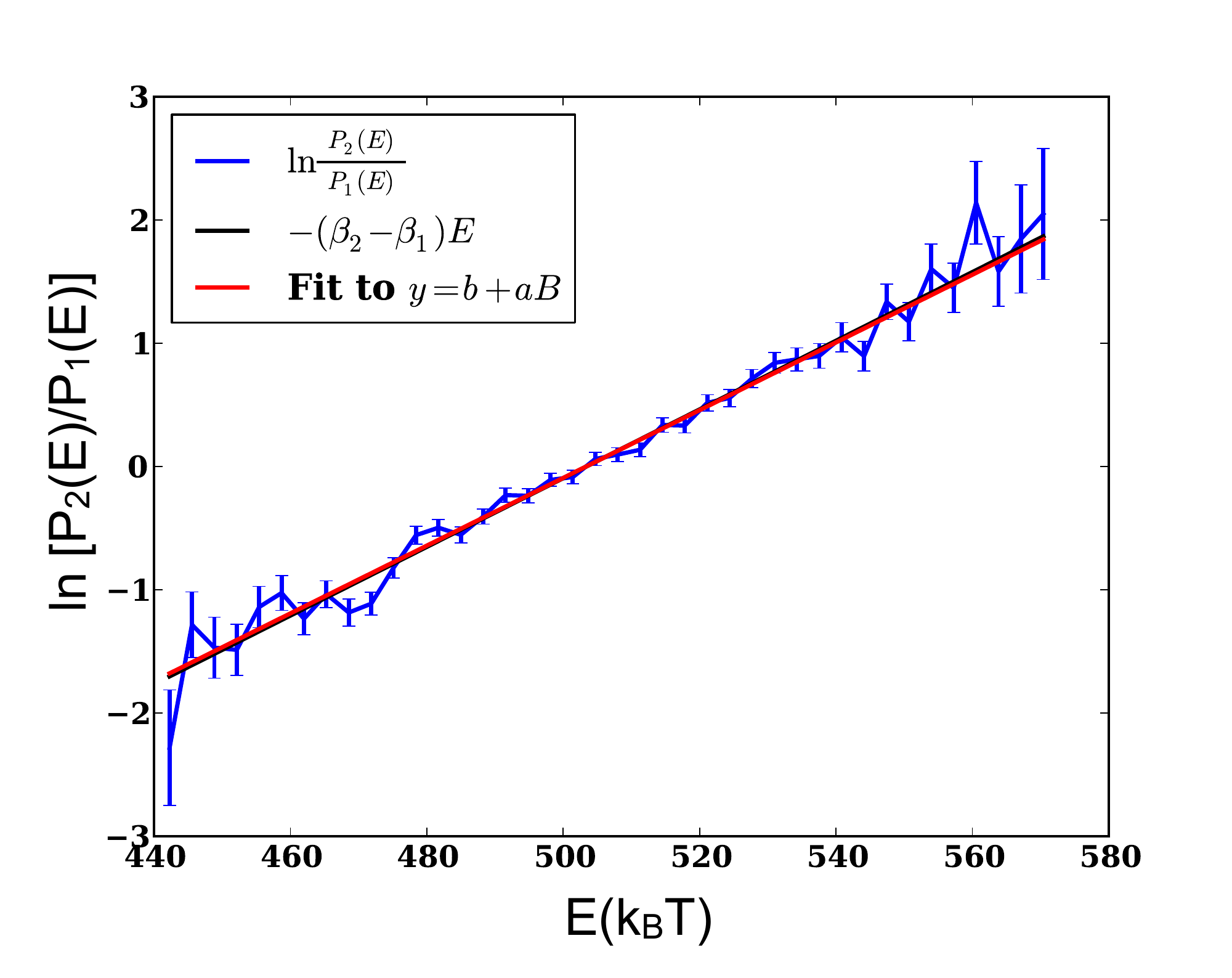} \\
a & b 
\end{tabular}
\caption{{\bf Differences in validation of Berendsen and
    Nos\'{e}-Hoover thermostats.} Berendsen temperature control
  produces simulations deviating greatly from the true distribution;
  in this case, the slope $\beta_2 - \beta_1$ of the kinetic energy
  log ratio is 7 times higher than it should be, 68 standard
  deviations away from the true value.  The Nos\'{e}-Hoover
  thermostat, like most others examined here, gives a slope
  statistically indistinguishable from the proper slope for the
  kinetic energy portion of the canonical
  ensemble.\label{fig:compberendsen}}
\end{figure}

This comparison is presented in
Table~\ref{table:thermostat_validation}, with all estimates and errors
computed using maximum likelihood methods described in this paper.  We
see that all temperature control methods appear to be consistent with
a canonical ensemble, with the exception of the Berendsen temperature
control method, with deviations from the true slope generally 1
$\sigma$ or less. NVE kinetic energy distributions deviate from the
canonical ensemble, though interestingly, potential energy
distributions do not deviate from the correct distribution to a
statistically noticeable level.  In all cases where there are
deviations of the kinetic energy, the distributions of the potential
energies are closer to the true distribution than the kinetic energy
or total energy distributions are; as noted, for NVE, the potential
energy distribution is statistically indistinguishable from the NVT
potential energy distribution.

\subsubsection{Molecular example: the effect of large step size}
It is well known that step sizes that are too large can lead to rapid
heating of a NVE molecular dynamics simulation as the integration
deviates from the conserved energy trajectory.  This deviation was one
of the initial motivations leading to the development of thermostats.
However, using a thermostat to bleed out the extra thermal energy
created by violations of the conservation of energy effectively
creates a steady state system. The system has both heat being both
pumped in by numerical integration error and pumped out by the
thermostat, with the the average kinetic energy having the desired
average.  However, this steady state process does not necessarily have
the correct Boltzmann probability distribution.

There has been relatively little investigation of the effect of step
size on the ensemble itself when temperature control is
applied,~\cite{Pastor_1988} especially for atomistic simulations.
Here, we examine step sizes from 8 fs to 40 fs.  In the Gromacs code,
a step size of 48 fs with Lennard-Jones argon cause segmentation
faults within just a few ns and therefore represents the upper limit
of stability with a thermostat coupling constant with $\tau_T = 1$ ps.
In these units, the reduced time is $\sigma(M/\epsilon)^{1/2}=0.1245$
picoseconds, so the stability limit is about 0.386 reduced time units.

However, being below the limit of stability does not necessarily mean
that the ensemble is correctly reproduced.  To analyze the
distributions generated by long step sizes, we use the
Bussi-Parrinello thermostat algorithm, and step sizes ranging from 8
fs to 40 fs (Table~\ref{table:stepsize}). Uncorrelated potential energy
samples were 20 ps apart as determined by the {\tt timeseries} module,
consistent over all steps sizes to within 10\%.  Uncertainties in
effective temperature are determined directly from the subsampled
kinetic energies, rather than using Gromacs {\tt g\_energy} output, in
order to have a more consistent treatment of uncertainties between
different observables.

\begin{table}[tbp]
\resizebox{16.5cm}{!}{
\begin{tabular}{|ccc|cc|cc|cc|}
\hline
 $\Delta t\;$(fs) & \multicolumn{2}{c}{True $T_{\mathrm{low}}=132.915$ K}   &\multicolumn{6}{c|}{True $\Delta T = 4.223$ K} \\
\hline
                & $T_{\mathrm{low}}$(K) & $\sigma$ deviation &\multicolumn{3}{c}{Estimated $\Delta T$}  & \multicolumn{3}{c|}{$\sigma$ deviation slope from true} \\ 
\hline 
                &               &                     &\multicolumn{2}{c}{total} &   \multicolumn{2}{c}{potential}   &  \multicolumn{2}{c|}{kinetic}      \\
\hline
       8        &  132.924 $\pm$ 0.040  &  0.2  & 4.230 $\pm$ 0.047 &  0.2 &  4.237 $\pm$  0.058 &  0.2 &  4.186 $\pm$ 0.063 & 0.6  \\
16              &  132.933 $\pm$ 0.028  &  0.7  & 4.183 $\pm$ 0.032 &  1.2 &  4.253 $\pm$  0.040 &  0.8 &  4.106 $\pm$ 0.043 & 2.7  \\
24              &  132.933 $\pm$ 0.023  &  0.8  & 4.058 $\pm$ 0.026 &  6.4 &  4.140 $\pm$  0.032 &  2.6 &  4.023 $\pm$ 0.035 & 5.8  \\
32              &  132.905 $\pm$ 0.020  &  0.5  & 3.967 $\pm$ 0.030 &  8.6 &  4.199 $\pm$  0.028 &  0.9 &  4.054 $\pm$ 0.022 & 7.6  \\
40              &  132.948 $\pm$ 0.019  &  1.8  & 3.988 $\pm$ 0.020 & 11.6 &  4.178 $\pm$  0.026 &  1.7 &  3.877 $\pm$ 0.027 & 12.9 \\
40 ($E_{\mathrm{kin}}$ ave)   &  132.917 $\pm$ 0.018  &  0.1  & 4.266 $\pm$ 0.021 &  2.6 &  4.275 $\pm$  0.026 &  2.0 &  4.296 $\pm$ 0.029 & 2.6  \\
\hline
\end{tabular}
}
\caption{{\bf Effect of step size on ensemble consistency.} Total and
  kinetic energy gradually deviate from the true ensemble as step size
  increases, becoming statistically noticeably near, but not at the
  instability point.  Potential energy distributions deviate less
  significantly from a canonical distribution than the kinetic energy
  distributions. The average half step kinetic energy estimator using
  the leapfrog Verlet deviates less from the true
  distribution.\label{table:stepsize}}
\end{table}

In Table~\ref{table:stepsize}, we note that total and kinetic energy
gradually deviate from the true ensemble with the deviation becoming
extremely large near the instability point. For this particular
system, average temperatures determined by averages of the kinetic
energy from a simulation (shown for the lower temperature simulation
in Table~\ref{table:stepsize}) are not as useful to distinguish
systems that are being forced back to the desired average kinetic
energy using the thermostat.

Interestingly, potential energy distributions deviate much less
significantly from the canonical distribution than the kinetic energy
distributions to the extent that is this deviation is not
statistically significant.  This may relate to the fact that standard
estimator of the kinetic energy in the velocity Verlet algorithm, the
sum of the squared full step velocities times the masses, is not as
accurate as the estimator of the kinetic energy of the leapfrog Verlet
algorithm, which uses the averaged half-step kinetic
energies. Although deviations increase with the square of the step
size in both cases, the full step kinetic energies deviate more
quickly.~\cite{Cuendat_2007} We note that it appears to be the choice
of kinetic energy estimator, not the integration method per se that
makes a difference, since the two methods give identical NVE
trajectories up to numerical precision.  The hypothesis that the
choice of kinetic energy estimator may make a difference was confirmed
by performing the same 40 fs time step simulation with the leapfrog
Verlet integrator and the Bussi-Parrinello algorithm, resulting in
significantly better kinetic energy distribution without statistically
altering the potential energy distributions.
We note that in this case, although the deviation
is statistically very clear, it is not necessarily that large.  Even
for the 40 fs step kinetic energy, the fitted temperature difference
is only off 10\%, which is about 0.4 K, which will not make a
difference for most applications. We also note that simulations of
different molecular systems with different potential functions may
have different deviations from ensemble consistency as a function of
distance from the time step stability point.

\subsubsection{Example: Examining the effect of cutoffs on ensemble consistency} 

An abrupt cutoff of a radial potential function creates a
discontinuity in the force, resulting in steadily increasing
temperature in an NVE simulation.  This temperature rise can, as in
the case of large time step, again be disguised by adding a
thermostat, creating a steady state system that does not necessarily
obey the canonical distribution.  We can examine the effect of this
truncated potential on the NVT ensemble using our ensemble consistency
tests.  We run the same Lennard-Jones argon system with abrupt cutoffs
at $r_c = 2.0\sigma$, $2.5\sigma$, $3.0\sigma$, $3.5\sigma$, and
$4.0\sigma$, where $\sigma$ here is the Lennard-Jones size, not the
standard deviation.  Because of quirks in the way Gromacs handles
abrupt cutoffs, we create an abrupt cutoff using a potential switch
over a distance of $10^{-9}$ nm, which on the integration time scale
effectively becomes a discrete cutoff.  We can measure how much such a
simulation violates conservation of energy by monitoring the average
increase in the conserved quantity per unit time.  In these
simulations, we use the Bussi-Parrinello thermostat, with $\tau_T =
1.0$ ps, approximately 120 times the time step, with $T_{\mathrm{low}}
= 132.915$ and $T_{\mathrm{high}} = 137.138$.  We can measure the
magnitude of energy drift by monitoring the change in the conserved
quantity over time which varies from $9.40\times10^3$ kJ mol$^{-1}$
ns$^{-1}$ for $r_c = 2.0\sigma$ to 78 kJ mol$^{-1}$ns$^{-1}$ for $r_c =
4.0\sigma$.  Times between uncorrelated samples, as determined by
potential energy differences, were no larger than 25 ps for all
systems, so we use this sampling time frequency all three quantities.

\begin{table}[tbp]
\resizebox{16.5cm}{!}{
\begin{tabular}{|cccc|cc|cc|cc|}
\hline 
         &                                &   T$_{\mathrm{low}}$ = 132.915        &  & \multicolumn{6}{c|}{True $\Delta T = 4.223$} \\
  r$_c$ (LJ $\sigma$) &  E$_{\mathrm{cons}}$ (gained kJ/ns) & Estimated T$_{\mathrm{low}}$ (K) & $\sigma$ deviation  &\multicolumn{3}{c}{Estimated $\Delta T$} & \multicolumn{3}{c|}{$\sigma$ deviation} \\
\hline
         &                                &                  &  &  \multicolumn{2}{c}{total} & \multicolumn{2}{c}{potential} & \multicolumn{2}{c|}{kinetic}       \\
\hline
       2 &         9400                  & 133.952 $\pm$ 0.045 & 23.0  &  4.102 $\pm$ 0.058 &  2.1 & 4.018 $\pm$ 0.084 & 2.4 & 4.122 $\pm$ 0.070 & 1.4 \\
     2.5 &         1140                  & 133.043 $\pm$ 0.045 & 2.9   &  4.206 $\pm$ 0.052 &  0.3 & 4.177 $\pm$ 0.059 & 0.8 & 4.176 $\pm$ 0.071 & 0.7 \\
       3 &          239                  & 132.941 $\pm$ 0.045 & 0.6   &  4.232 $\pm$ 0.051 &  0.2 & 4.291 $\pm$ 0.065 & 1.1 & 4.213 $\pm$ 0.071 & 0.1 \\
     3.5 &          104                  & 132.930 $\pm$ 0.045 & 0.3   &  4.226 $\pm$ 0.050 &  0.1 & 4.302 $\pm$ 0.058 & 1.4 & 4.135 $\pm$ 0.070 & 1.2 \\
       4 &           78                  & 132.929 $\pm$ 0.045 & 0.3   &  4.302 $\pm$ 0.050 &  1.6 & 4.307 $\pm$ 0.057 & 1.6 & 4.192 $\pm$ 0.071 & 0.4 \\
\hline
\end{tabular}
}
\caption{{\bf Effect of abrupt cutoff on ensemble validation.}
  Distributions are surprisingly ensemble consistent for most values
  of abrupt cutoff for Lennard-Jones spheres, with only the shortest
  cutoff distances (less than 3 LJ $\sigma$) showing statistically clear
  violations.~\label{table:cutoff}}
\end{table}

We see in Table~\ref{table:cutoff} that the distributions are
surprisingly ensemble consistent for most values of abrupt cutoff for
Lennard-Jones spheres despite the fact that the simulation is gaining
more than 200 kJ/mol/ns with a 3 $\sigma$ cutoff.  We note that in
this case, the deviation from desired temperature as calculated from
average kinetic energy is fairly clear (23 standard deviations for a 2
$\sigma$ cutoff!) and therefore this measure appears to be better at
distinguishing violations from the correct distribution than the
ensemble consistency check. This contrasts with the case of varying
step size, where the ensemble consistency check was more sensitive
than the deviation from the correct average kinetic energy.  Clearly,
multiple validation methods should always be performed!
\subsubsection{Validating the gap selection criteria for molecular systems}

Finally, we attempt to validate our rule of thumb for the ideal
temperature gap with molecular systems, since it was derived for a
simplified model system. We test the ability to detect error using the
same Lennard-Jones argon system with time step $\Delta t = 32$ fs
using velocity Verlet, as for higher temperatures, a time step of
$\Delta t = 40$ fs can crash in simulations extending for hundreds of
ns long. Measuring the heat capacity as 8.5 kJ mol$^{-1}$
K$^{-1}$ at 135 K leads to a standard deviation of 36 kJ/mol and an
estimated ideal temperature gap of 6 K between the means of the two
total energy distributions.  We see
(Fig.~\ref{table:molecular-temperature-gap}) that we are most
sensitive to error in the total energy between 1 and 2 times the
estimated gap, meaning that our analytical guidelines were close, but
that a slightly larger gap might sometimes be more effective in
identifying errors.  We note that the kinetic energy standard
deviation at 135 K (24 kJ/mol) is only about 2/3 of the total energy
standard deviation, but since the total heat capacity (8.5 kJ K$^{-1}$
mol$^{-1}$) is more than twice as large as the ideal gas heat capacity
(3.72 kJ K$^{-1}$ mol$^{-1}$ for this size of system), the kinetic
energy distributions have closer mean values than the total energy
distributions. Thus the range of peak discrimination for kinetic
energy still falls in the 1 to 2 times the ``twice the central
standard deviation'' rule of thumb, when using the distribution of
kinetic energies.  For molecular systems, the ideal gap might
therefore be better estimated using a temperature gap 1.5 to 2 times
the estimated cap range.  However, a relatively wide range of values
allows discriminating lack of ensemble validity if sufficient data is
collected.

\begin{table}[tbp]
\resizebox{16cm}{!}{
\begin{tabular}{|ccccccccc|}
\hline
            &           &   &     \multicolumn{2}{c}{Kinetic} &  \multicolumn{2}{c}{Potential} & \multicolumn{2}{c|}{Total}  \\
$\Delta T/T$  & $\beta_2-\beta_1$ & n$\times$gap$_{\mathrm{opt}}$   & Est. slope & $\sigma$ deviation & Est. slope & $\sigma$ deviation & Est. slope & $\sigma$ deviation \\
0.0156 &   0.01393 & 0.4  & 0.01302 $\pm$ 0.00023  &  4.0 & 0.01413 $\pm$ 0.00021 & 0.9  &  0.013518 $\pm$ 0.00016 & 2.6  \\ 
0.0313 &   0.02787 & 0.7  & 0.02598 $\pm$ 0.00025  &  7.7 & 0.02585 $\pm$ 0.00018 & 1.4  &  0.026433 $\pm$ 0.00018 & 7.8  \\ 
0.0469 &   0.04182 & 1.1  & 0.03892 $\pm$ 0.00027  & 10.7 & 0.04152 $\pm$ 0.00027 & 1.1  &  0.039990 $\pm$ 0.00023 & 8.1  \\ 
0.0626 &   0.05578 & 1.4  & 0.05190 $\pm$ 0.00031  & 12.6 & 0.05526 $\pm$ 0.00031 & 1.8  &  0.053343 $\pm$ 0.00029 & 8.6  \\
0.0938 &   0.08378 & 2.1  & 0.07791 $\pm$ 0.00042  & 14.2 & 0.08290 $\pm$ 0.00045 & 2.0  &  0.079531 $\pm$ 0.00049 & 8.6  \\
0.1251 &   0.11189 & 2.8  & 0.10464 $\pm$ 0.00059  & 12.2 & 0.11150 $\pm$ 0.00071 & 1.6  &  0.107768 $\pm$ 0.0010  & 4.2  \\ 
0.1877 &   0.16867 & 4.2  & 0.1522  $\pm$ 0.0016   & 10.1 & 0.1654  $\pm$ 0.0022  & 1.5  &  0.165805 $\pm$ 0.0057  & 0.5  \\
0.2502 &   0.22645 & 5.6  & 0.2099  $\pm$ 0.0037   &  4.5 & 0.220   $\pm$ 0.010   & 0.6  &  0.249    $\pm$ 0.072   & 0.3  \\
\hline
\end{tabular}
}
\caption{{\bf Molecular validation of ideal gap guidelines}. We test
  the temperature gap for maximum discrimination rule with
  Lennard-Jones argon with time step $\Delta t = 32$ fs. Maximum
  discrimination of error in the ensemble consistency for the
  different energy terms occurs between 1 and 2 times the estimated
  gap rule (column 3).\label{table:molecular-temperature-gap}}
\end{table}

\subsection{Examining pressure control algorithms}

There are currently three pressure control algorithms implemented in
Gromacs: Berendsen,~\cite{Berendsen1984}
Parrinello-Rahman,~\cite{Parrinello81,Nose83} and the
Martyna-Tuckerman-Tobias-Klein (MTTK)
algorithm.~\cite{Tuckerman2006,Martyna1996} The first two are defined
using the leapfrog integrator in Gromacs, and the first and last are
defined using the velocity Verlet integrator.  We next examine the
same small argon system for fluctuations of enthalpy and volume, and
the joint fluctuation of volume and energy.  A velocity Verlet
integrator was used except for Parrinello-Rahman, with $\Delta t = 8$
fs.  We set the pressure coupling $\tau_p$ to 5 ps in all cases and
use $P=90$ bar and $T=125$ K as the average pressure and temperature,
resulting in a system well below the critical point.  When testing
volume fluctuations or joint energy and volume fluctuations, a low
pressure of 30 bar and a high pressure of 150 bar were used ($\Delta P
= 120$ bar), except for Berendsen pressure control, where low and high
pressures of 88 and 92 bar ($\Delta P = 4$ bar) were used.  A lower
range is needed for the Berendsen weak coupling algorithm as the
volume distributions are far smaller than is correct for the
distribution (already demonstrating a problem).  When testing enthalpy
fluctuations or joint energy and volume fluctuations, a low
temperature of 121.431 K and a high temperature of 128.569 K ($\Delta
\beta = 0.054987~k_{B}T^{-1}$, $\Delta T = 7.138~$K) were used,
generated using an estimated $C_P$ of 10.2 kJ/mol from short initial
simulations for this system using the estimated gap formula.  For
Berendsen thermostat simulations, a temperature range of 124.108 to
125.892 K was used ($\Delta \beta = 0.013736$, equivalent to $\Delta T
= 1.784$), as again the overlap between the distributions is very
poor for wider parameter differences. Nose-Hoover temperature control
with $\tau_T = 1$ ps was used for both Parrinello-Rahman and MTTK
algorithms.  The $\Delta P$ for joint energy and volume comparisons is
smaller because the simulations are run at different temperatures and
is equal to $\Delta P = 114.861$ bar for Parrinello-Rahman and MTTK
and 2.715 bar for Berendsen.

Looking at Table~\ref{table:barostat_validation}, we see that the
Parrinello-Rahman and MTTK algorithms reproduce very accurately the
correct enthalpy distributions, deviating very little from the correct
$\Delta \beta$, with very high statistical confidence.  The precision
is high partly because the time between uncorrelated samples (in this
case, determined from the largest correlation time of either the
energy or the volume) is quite short, in the range of 4-6 ps.  The
volume distributions, however, are somewhat off, with the effective
$\Delta P$ in both cases near 115 $\pm$ 0.6 instead of 120.  For most
cases, this will be sufficiently accurate to model physical processes
(and is far better than the Berendsen results), but might not be
sufficiently accurate for very high precision thermodynamic
measurements.  The $9 \sigma$ deviation from the true answer is again
not necessarily a sign of how bad the simulation is.  In this case,
because the slope is nearly correct, it sign that it is statistically
very {\em likely} the simulation is at least somewhat off rather than
simply being very {\em bad}.  Similar patterns are seen in the joint
distribution of $E$ and $V$, where the effective $\Delta P$ is still
off by about 5 bar (or around 5\%). The deviations are similar for
both MTTK and Parrinello-Rahman, even though these integration
routines are mostly separated in the Gromacs code.

For Berendsen, the results are uniformly bad.  In all cases, the
deviation from the expected values is significantly higher than with
MTTK or Parrinello-Rahman, with the slopes being much further from the
true value even though the statistical error is much higher as well.
This deviation exists even though the average temperatures and
pressures in the Berendsen case were all well within statistical
noise. For example, for the joint distribution analysis, the low and
high average pressures were indeed $87.996 \pm 0.003$ and $91.998 \pm
0.005$ bar and the average temperatures were $125.865 \pm 0.015$ K and
$124.081 \pm 0.015$, well within the statistical noise.  Errors in the
fitting parameters are therefore due to unphysically narrow
distributions, not the average values themselves.  We note one other
potential strange problem with Berendsen volume control combined with
Bussi-Parrinello thermostat.  The autocorrelation times are much
longer than with other simulation variables, on the order of 20 ps for
the energies and 110-130 ps for the volumes.  It is not clear what
exactly is causing such slow change of these variables when the time
constants themselves are much lower---in this case $\tau_T = 1$ ps and
$\tau_P = 5$ ps---but perhaps indicates another reason to avoid
Berendsen pressure control.
 
\begin{table}[tbp]
\resizebox{16.5cm}{!}{
\begin{tabular}{|l|cc|cc|cccc|}
\hline
                     & \multicolumn{2}{c}{enthalpy} &   \multicolumn{2}{c}{volume} & \multicolumn{4}{c|}{joint energy and volume} \\
\hline
Barostat              & $\Delta$ slope & $\sigma$ deviation & $\Delta$ slope & $\sigma$ deviation & $\Delta$ slope & $\sigma$ deviation &  $\Delta$ slope & $\sigma$ deviation \\
\hline
       Berendsen  &  4.176 $\pm$ 0.121 & 19.8  &  79.5 $\pm$ 4.4    &   17.1  &  0.69 $\pm$ 0.14   & 7.6 & -318.661 $\pm$ 7.322 & 43.9 \\
Parrinello-Rahman &  7.022 $\pm$ 0.033 &  3.5  & 114.58 $\pm$ 0.57  &    9.5  &  7.168 $\pm$ 0.036 & 0.8 & 110.971 $\pm$ 0.529  & 7.4 \\
     MTTK         &  7.105 $\pm$ 0.029 &  1.2  & 115.51 $\pm$ 0.50  &    9.0  &  7.152 $\pm$ 0.031 & 0.5 & 111.312 $\pm$ 0.457  & 7.8 \\
\hline
\end{tabular}
}
\caption{{\bf Ensemble validation of pressure control algorithms}
  Tests of enthalpy distribution, volume distribution, and joint
  energy and volume distributions.  The Berendsen barostat fails badly in
  all three tests. The other two barostat give correct enthalpy
  distributions, but have small ($\Delta P$ off by 5 bar or $\approx 5$\%) but
  statistically clear (7--9$\sigma$) errors in the volume
  distributions.~\label{table:barostat_validation} }
\end{table}

\subsection{Water simulations}
We also examine a somewhat more typical system for molecular
simulation, a small box of 900 TIP3P water, a size that might be used
to compute pure water properties or small molecule solvation free
energies.  We again use velocity Verlet integration (with the
exception of the Gromacs stochastic integration, which is only defined
for the leapfrog Verlet algorithm) with linear center of mass momentum
removal every step and a long range homogeneous dispersion correction
applied to the energy and virial. We use a Lennard-Jones switch
between 0.8 and 0.9 nm with a neighborlist at 1.0 nm and particle
mesh Ewald electrostatics with cutoff 1.0 nm, PME order 6, and Ewald
cutoff tolerance of $10^{-6}$.  In all cases, neighborlist update
frequency of 10 steps was used with a time step of 2
fs. SETTLE~\cite{rattle,Miyamoto1992} was used to constrain the water
bonds and angles, and a total of 10 million steps (20 ns) were
simulated, with the last 19 ns used for analysis.  Temperature
coupling algorithms were carried out with a coupling constant of
$\tau_T = 1.0$~ps for the NVT simulations and $\tau_T=5.0$ and
$\tau_P=5.0$ for the NPT simulations. The low temperature is 298 K and
301 K, with $\Delta T/T = 0.01$ estimated from $\sigma_E$ in the total
energy from a single short simulation using the relationships for the
ideal temperature gap. For the NPT simulations, using a $\sigma_V =
0.25 \mathrm{nm}^3$ at 1 bar from a short simulation predicts a
$\Delta P$ of 238 bar using the formula presented here, but to err on
the side of having sufficient samples we instead use $\Delta P = 175$
with the low pressure at 1 bar and the high at 351 bar, though we are
potentially losing some precision.  In the case of Berendsen pressure
control, we used $\Delta T = 1$K, and $\Delta P = 30$ bar to ensure
overlap. For NVT, the interval between uncorrelated samples is
determined from correlation times of the potential energy which is 2
ps for all methods except the Andersen method, where we use 4 ps.  For
NPT, we use the maximum of the uncorrelated sample intervals between
the volume and the energy.  Correlation times for MTTK are much
smaller, around 0.3-0.4 ps for both energy and volume, whereas for
Berendsen the energy and volume uncorrelated sample intervals are both
4 ps, and for Parrinello-Rahman, the energy and volume intervals are 6
ps and 0.4 ps, respectively.  Thus, the NPT MTTK results are somewhat
more precise.


\begin{table}[tbp]
\resizebox{16.5cm}{!}{
\begin{tabular}{|l|cc|cc|cc|}
\hline
                     & \multicolumn{2}{c}{total} &   \multicolumn{2}{c}{potential} &  \multicolumn{2}{c|}{kinetic} \\
\hline
Thermostat                     & Slope & $\sigma$ deviation & Slope & $\sigma$ deviation & Slope & $\sigma$ deviation \\ 
\hline
         Berendsen &  51.6 $\pm$ 1.1  & 44.2  &   7.20 $\pm$ 0.12 & 34.7 &  4.86 $\pm$ 0.12  & 15.8  \\
        Stochastic &   2.998 $\pm$ 0.059  &  0.04 &   2.944 $\pm$ 0.069 &  0.8 &  3.032 $\pm$ 0.090  &  0.4  \\
   Nos\'{e}-Hoover &   2.921 $\pm$ 0.058  &  1.4  &   2.953 $\pm$ 0.068 &  0.7 &  2.837 $\pm$ 0.089  &  1.8  \\
          Andersen &   3.028 $\pm$ 0.083  &  0.4  &   3.114 $\pm$ 0.098 &  1.2 &  2.870 $\pm$ 0.126  &  1.0  \\
Andersen (Massive) &   3.086 $\pm$ 0.083  &  1.0  &   3.048 $\pm$ 0.097 &  0.5 &  3.136 $\pm$ 0.127  &  1.0  \\ 
  Bussi-Parrinello &   2.955 $\pm$ 0.058  &  0.8  &   2.956 $\pm$ 0.068 &  0.6 &  3.021 $\pm$ 0.090  &  0.2  \\
\hline
\end{tabular}
}
\caption{{\bf Ensemble validation of different temperature control
    algorithms with water} $\Delta T = 3$ K corresponding to a inverse
  temperature slope of is 0.004023 $(k_B T)^{-1}$. Results are consistent
  with those performed with argon, with all temperature control
  algorithms ensemble consistent except for Berendsen.\label{table:water_validation_thermostat} }
\end{table}
We first examine the NVT results in
Table~\ref{table:water_validation_thermostat}. These results are
completely in keeping with the argon results before, with all
temperature control methods well within statistical error, with the
exception of Berendsen, which is again wildly incorrect.  These
results demonstrate that the utility of ensemble validation is
applicable to more typical molecular simulations, with data set sizes
that are more typical for a standard testing pipeline.

\begin{table}[tbp]
\resizebox{16.5cm}{!}{
\begin{tabular}{|l|cc|cc|cccc|}
\hline
                     & \multicolumn{2}{c}{enthalpy} &   \multicolumn{2}{c}{volume} & \multicolumn{4}{c|}{joint energy and volume} \\
\hline
Barostat              & Slope & $\sigma$ deviation & Slope & $\sigma$ deviation & Slope & $\sigma$ deviation &  Slope & $\sigma$ deviation \\
\hline
Berendsen             & 1.03 $\pm$ 0.15 & 0.2 &  262  $\pm$ 25        & 9.0  &  1.67 $\pm$ 0.21 & 3.3 & 250 $\pm$ 30 & 7.4 \\
Parrinello-Rahman     & 2.65 $\pm$ 0.21 & 1.7 &  309.3 $\pm$ 3.7  & 11.1 &  4.09 $\pm$ 0.34 & 3.2 & 354 $\pm$ 19 & 0.2 \\
MTTK                  & 2.978 $\pm$ 0.053 & 0.4 &  335.7 $\pm$ 3.9  & 3.7  &  3.026 $\pm$ 0.074 & 0.4 & 345.7 $\pm$ 4.6  & 0.6  \\

\hline
\end{tabular}
}
\caption{{\bf Ensemble validation of pressure control algorithms in
    water.} Tests of enthalpy distribution, volume distribution, and
  joint energy and volume distributions.  For Parrinello-Rahman and
  MTTK, the true $\Delta T = 3$ and true $\Delta P = 350$, while for
  Berendsen, they are $\Delta T = 1$ and $\Delta P = 30$ in the joint
  energy and volume case.  The Berendsen barostat performs
  significantly worse, requiring a much narrower range of variables to
  get any overlap.  The other two barostat give statistically valid
  enthalpy distributions, with MTTK appearing to have fairly accurate
  volume distributions and with Parrinello-Rahman having somewhat
  worse volume behavior.~\label{table:water_barostat_validation} }
\end{table}

From the NPT results in Table~\ref{table:water_barostat_validation},
we see Parrinello-Rahman and MTTK have reasonable performance in
describing the enthalpy distribution.  Berendsen in this case is also
reasonable, perhaps because the entropy contribution dominates for the
nearly incompressible water. MTTK has somewhat better results for
volume fluctuations than Parrinello-Rahman.  It is interesting to
speculate on exactly the source of the difference between the volume
fluctuation results in the argon and the water examples.  In the argon
example, both pressure control algorithms had small but statistically
noticeable errors that were consistent between the two algorithms. In
the water example MTTK appears to be fairly ensemble consistent,
whereas Parrinello-Rahman is slightly worse.  Parrinello-Rahman with
leapfrog is known to be inexact because the pressure lags by one time
step, as the pressure and temperature are not both known at a given
time $t$ until after the next half step. This may be more of a problem
in the case of water because with a higher compressibility, volume
integration is a stiffer equation, requiring more exact solutions.  We
can tentatively conclude that typical aqueous simulations using MTTK
may be more consistent with an NPT ensemble than Parrinello-Rahman,
though both are far better than Berendsen temperature control.

\section{Tools}
To make these ensemble consistency checks easier, we have created a
set of tools to assist other researchers to more easily measure the
ensemble validations. This code is hosted by SimTK, at {\tt
  http://simtk.org/home/checkensemble} and includes automatic plotting
of linear and nonlinear graphs, linear, nonlinear, and maximum
likelihood parameter analysis.  These software tools were used for all
analysis in this paper.  These tools include example code for parsing
Gromacs, CHARMM and Desmond output files for ensemble consistency for
both NVT and NPT simulations, including testing enthalpy, volume, and
joint energy and volume fluctuations. Scripts to regenerate all the
harmonic oscillator analytic tests described in this paper are also
included in the distribution.

\section{Conclusions}

We have shown that for molecular distributions characterized by
Boltzmann distributions, which is true for essentially all molecular
simulations performed at NVT and NPT, we can easily check for
consistency with the intended ensemble regardless of the details of
the simulation.  We simply require pairs of simulations with differing
external parameters such as temperature, pressure, or chemical
potential.  These paired simulations allow system-dependent properties
such as densities of states to cancel out, resulting in a linear
relationship between the distribution of extensive quantities such as
energy, volume, enthalpy, and number of particles.  Importantly, the
constant of proportionality in this linear relationship is completely
determined by the intensive variables that are set by the user.

Tests of simple model systems shows that these relationships are not
only qualitatively useful but also that with proper error analysis can
provide quantitative validation of the statistics of the
distributions.  We have demonstrated the utility of these
relationships with simple analytical toy models of harmonic
oscillators in both the NVT and NPT ensembles as well as with
molecular simulations of argon and water.  We see that these ensemble
consistency relationships are able to identify thermostats and
barostats that are inconsistent with the ensemble as well as identify
differences in distributions caused by long time steps or abrupt
cutoffs. All tested thermostats except the Berendsen thermostat give
statistically good results.  Barostats were somewhat more problematic,
with MTTK giving the best results and Parrinello-Rahman being
acceptable for many uses, while Berendsen pressure control is simply
wrong for any calculation where volume fluctuations are important.  In
all cases, simpler checks such as making sure estimators of
quantities like the temperature and pressure calculated from the
kinetic energy and the virial do indeed have the correct value are
useful as diagnosis tools, and may occasionally identify problems that
are not easily identified by the ensemble consistency methods tested
here.

These relationships between pair distributions are true for all
differences in applied external thermodynamic variables.  However,
there are statistical reasons for choosing specific differences in the
parameters.  We have shown that for simple potentials both small and
large differences in the applied system parameters lead to difficulty
in distinguishing systems with errors from systems with the correct
distributions.  We have also shown that for typical probability
distributions, choosing distributions whose means are separated by
gaps two to four times the sum of the standard deviations appears to
maximize the ability to discriminate between data that is or is not
consistent with the desired ensemble, erring on the shorter side in
cases where less data might be available.  It is also important not to
underestimate the autocorrelation time for the energy variables to be
able to accurately use the error estimates, as it may give
inaccurately high deviations from the correct distribution.  Indeed,
in typical simulation cases, the ability to properly estimate
correlation times may be the largest source of uncertainty, as all
other parts of calculations are highly robust.  We also emphasize that
the size of the statistical deviation is a measure of how certain we
are of the discrepancy, not necessarily the size of the discrepancy,
as with sufficient data, we can statistically identify with a high
certainty small deviations that generally do not affect simulation
properties significantly.  Finally, we note that these are very
sensitive necessary tests, but they are not sufficient tests; they
cannot guarantee that all states with the same energy are equally
sampled, nor can they guarantee that all important regions of phase
space are sampled.

We have also developed simple software tools to easily perform the
statistical validation discussed here, requiring only lists of the
relevant extensive variables and specification of the intensive
applied variables.  These tools can be easily incorporated into the
workflow for molecular simulation testing, hopefully greatly reducing
the difficulty of determining whether a given algorithm or software
program is producing the desired thermodynamic ensemble.  Future
potential improvements of these tools include adapting the tools for
grand canonical simulations and translating the relatively
unsophisticated quantile validation into full statistical hypothesis
testing.


\acknowledgement The author wishes to thank Ed Maginn (Notre Dame
University) and Lev Gelb (UT-Dallas) for comments and suggestions, Joe
Basconi (University of Virginia), Daniel Sindhikara (Institute for
Molecular Sciences, Japan), and John Chodera (UC-Berkeley) for careful
reading of the manuscript, and SimTK.org for hosting the code.

\begin{appendix}
\section{Grand Canonical Ensemble}
\label{appendix:grand}
Although no grand canonical simulations were carried out in this
study, all the equations are essentially equivalent in the case of the
isobaric-isothermal ensemble with $-\mu$ taking the place of $P$ and
$N$ taking the place of $V$.
\begin{eqnarray}
P(x,N|\beta,\mu) = \Xi(\beta,\mu)^{-1}\exp(-\beta E +\beta \mu N)
\end{eqnarray}
Examining the probability of $N$ at fixed $\beta$ and $P$ performed at
two different chemical potentials $\mu_1$ and $\mu_2$, we obtain
\begin{eqnarray}
P(x,N|\beta,\mu_1) &=& \Xi(\beta,\mu_1)^{-1}Q(\beta,N)\exp(\beta_1 \mu N) \nonumber \\
P(x,N|\beta,\mu_2) &=& \Xi(\beta,\mu_2)^{-1}Q(\beta,N)\exp(\beta_2 \mu N) \nonumber \\
\frac{P(N|\beta,\mu_2)}{P_1(N|\beta,\mu_1)} &=& \frac{\Xi(\beta,\mu_1)}{\Xi(\beta,\mu_2)}\exp([\beta \mu_1 - \beta \mu_2] N) \\
\ln \frac{P_2(N|\beta,\mu_2)}{P(N|\beta,\mu_1)} &=& \beta \left(-[(PV)_2 - (PV)_1] + [\mu_2 - \mu_1] N\right)
\end{eqnarray}
We note that in the grand canonical case, $N$ is already discrete, so
a histogramming approach introduces no additional approximations as
long as the histograms are fine grained down to integers.  For samples sizes large
enough that larger bins are required for accurate determination of
probabilities, the maximum likelihood method will be more accurate.

We can also treat the joint probability distributions of $N$ and $E$.
\begin{eqnarray}
P(N,E|\beta_1,\mu_1) &=& \Omega(N,E)\Xi(\beta_1,\mu_1)^{-1}\exp(-\beta_1 E+ \beta_1 \mu_1 N) \nonumber \\
P(N,E|\beta_2,\mu_2) &=& \Omega(N,E)\Xi(\beta_2,\mu_2)^{-1}\exp(-\beta_2 E+ \beta_2 \mu_2 N) \nonumber \\
\frac{P(N,E|\beta_2,\mu_2)}{P(N,E|\beta_1,\mu_1)} &=& \frac{\Xi(\beta_1,\mu_1)}{\Xi(\beta_2,\mu_2)}\exp(-[\beta_2-\beta_1]E + [\beta_2 \mu_2 - \beta_1 \mu_1] N) \\
\ln \frac{P(N,E|\beta_2,\mu_2)}{P(N,E|\beta_1,\mu_1)} &=& -(\beta_2(PV)_2 -\beta_1(PV)_1)-[\beta_2-\beta_1]E + [\beta_2 \mu_2 - \beta_1 \mu_1] N
\end{eqnarray}
This approach can easily be generalized to multiple chemical species,
especially when using maximum likelihood methods to allow minimization of the resulting multidimensional probability rations.   For example, for an arbitrary number of species $\vec{N}$ with associated chemical potentials $\vec{\mu}$ we have: 
\begin{eqnarray}
P(\vec{N},E|\beta_1) &=& \Omega(E,\vec{N})\Xi_1(\beta_1,\vec{\mu}_1)^{-1}\exp(-\beta_1 E+\beta_1 \vec{\mu}_1 \cdot \vec{N}) \nonumber \\
P(\vec{N},E|\beta_2) &=& \Omega(E,\vec{N})\Xi_2(\beta_2,\vec{\mu}_2)^{-1}\exp(-\beta_2 E+\beta_2 \vec{\mu}_2 \cdot \vec{N}) \nonumber \\
\frac{P(\vec{N},E|\beta_2,\mu_2)}{P(\vec{N},E|\beta_1,\mu_1)} &=& \frac{\Xi_1(\beta_1,\vec{\mu}_1)}{\Xi_2(\beta_2,\vec{\mu}_2)}\exp(-[\beta_2-\beta_1]E + \left[\beta_2 \vec{\mu}_2 -\beta_1\vec{\mu_1}\right]\cdot\vec{N}) \\
\ln \frac{P(\vec{N},E|\beta_2,\mu_2)}{P(\vec{N},E|\beta_1,\mu_1)} &=& -(\beta_2(PV)_2 -\beta_1(PV)_1)-[\beta_2-\beta_1]E + \left[\beta_2 \vec{\mu}_2 -\beta_1\vec{\mu_1}\right]\cdot\vec{N}
\end{eqnarray}

\section{Weighted least squares fitting to histogram ratios}
\label{appendix:error}
Assume we are collecting data from a continuous, one-dimensional
probability distribution in a histogram $H$ with $k=1 \ldots K$ bins.
We have $N$ total samples, with $\{n_1,n_2,\ldots,n_k\}$ observations
in each bin, so that $\sum_{k=1}^{K}n_k=N$.  The empirical probability of
finding an observation in bin $k$ is simply $p_k = n_k/N$.  Repeating
this experiment will lead to slightly different results for the
$p_k$. The standard estimator of variance of $p_k$ due to this sampling
variance is a standard result and is equal to $p_k (1-p_k)/N$.

Given two histograms $H_1$ and $H_2$ that have aligned bins with $N_1$
and $N_2$ samples each, the ratio of the probabilities of $H_2$ over
$H_1$ will be $r_k = p_{k,1}/p_{k,2}$ for each bin, where $p_{k,1}$
and $p_{k,2}$ are the probabilities in the $k$th bin for the first and
second simulation in the pair. The data in the two histograms are
collected independently, so the statistical variance in the logarithm
of the ratio $\ln r_k = \ln\left(p_{k,2}/p_{k,1}\right)$ will be to
first order:
\begin{eqnarray}
\Var{\ln r_k} &=& \frac{\Var{p_{k,1}}}{p_{{k,1}}^2} + \frac{\Var{p_{k,2}}}{p_{k,2}^2} \nonumber \\
              &=& \frac{1-p_{k,1}}{N_1 p_{k,1}} + \frac{1-p_{k,2}}{N_2 p_{k,2}} \nonumber \\ 
              &=& \frac{1}{n_{k,1}} - \frac{1}{N_1} + \frac{1}{n_{k,2}} - \frac{1}{N_2}
\end{eqnarray}
The variance in the ratio of the histograms themselves, useful for computing nonlinear estimates of the error will be: 
\begin{eqnarray}
\frac{\Var{r_k}}{r_k^2} &=& \frac{\Var{p_{k,1}}}{p_{{k,1}}^2} + \frac{\Var{p_{k,2}}}{p_{k,2}^2} \nonumber \\
\Var{r_k} &=& \left(\frac{p_{k,2}}{p_{k,1}}\right)^2\frac{\Var{p_{k,1}}}{p_{{k,1}}^2} + \frac{\Var{p_{k,2}}}{p_{k,2}^2} \nonumber \\
\Var{r_k} &=& \left(\frac{n_{k,2}N_1}{n_{k,1}N^2}\right)^2\left(\frac{1}{n_{k,1}} - \frac{1}{N_1} + \frac{1}{n_{k,2}} - \frac{1}{N_2}\right) 
\end{eqnarray}
Define a diagonal weight matrix $W$, with one over the variance in the
$i$th measurement along the diagonal. If we have a multivariate
function $F$ linearly dependent on data vector $X$ as $F = AY$, with
$A$ a constant matrix, then the covariance matrix of uncertainties
$\Cov{F}$ will be equal to $A~\Cov{Y}A^{T}$.  In the case of weighted
linear least squares, $\Cov{Y}$ is the matrix of weights $W$, where
$W_{ii} = \sigma^{-2}$, the variance of each histogram ratio
point. If $\alpha$ is the vector of parameters and $X$ is the
$(M+1)\times N$ matrix of observables, with the first column all ones
and the second through $(M+1)$th column the values of the observations
of the $M$ observables, then we will have for $\alpha$:
\begin{eqnarray*}
\alpha &=& (X^T W X)^{-1} X^{T} W Y 
\end{eqnarray*}
Plugging this into the equation for $\Cov{\alpha}$ in terms of
$\Cov{Y}$, some linear algebra leads to a covariance matrix of the
parameters $\vec{\alpha}$ of $(X^T W X)^{-1}$. If we have instead a
nonlinear least squares problem, at the minimum, we obtain a similar
covariance matrix, except that we replace $X$ with the Jacobian matrix
$J$ defined by $J_{ij} = \frac{\partial f(y_i,\vec{\alpha})}{\partial
  \alpha_j} $, where $f$ is the nonlinear model, and $y_i$ is the
$i$th data point. This leads to a final equation for the covariance of
the parameters:
\begin{eqnarray*}
\Cov{{\bf\alpha}} &=&  (J^T W J)^{-1}
\end{eqnarray*}
\section{Maximum Likelihood Estimation and Analytical Error Estimates}
\label{appendix:mle}
For a general Boltzmann-type probability distribution the ratio of probabilities must
satisfy:
\begin{eqnarray}
\ln \frac{P_2(\vec{X})}{P_1(\vec{X})} = \exp(-\vec{\alpha}\cdot\vec{X}) \label{eq:ratio.mle}
\end{eqnarray}
where the $\vec{X}$ are the $M$ sample variables (such as $E$ or $V$),
the $M+1$ $\alpha_j$ variables are the corresponding conjugate
variables specified by the simulation ensemble, and
$\vec{\alpha}\cdot\vec{X}_i$ is shorthand for $\alpha_0 +
\sum_{j=1}^M\alpha_j X_{j}$ rather than the standard dot product.

We develop the solution by finding maximum likelihood parameters along
the lines of the solution presented in (Ref.~\citep{Shirts2003}).  The
ratio in Eq.~\ref{eq:ratio.mle} can be interpreted as
$P(\vec{X}|1)/P(\vec{X}|2)$ where $P(\vec{X}|i)$ is the conditional
probability that an observation is from the $i$th simulation given
only the information $\vec{X}$. We would like to compute the
likelihood of a given set of $\alpha$ parameters given sets of
measurements with the specific simulation $i$ each set comes from
known.

Using the rules of conditional probabilities, and the fact that
$P(\vec{X}|1) + P(\vec{X}|2) = 1$,  we rewrite this probability distribution as
follows:
\begin{equation}
\frac{P(\vec{X}|2)}{P(\vec{X}|1)} =
\frac{\frac{P(2|\vec{X})P(\vec{X})}{P(2)}}{\frac{P(1|\vec{X})P(\vec{X})}{P(1)}} =
\frac{P(2|\vec{X})P(1)}{P(1|\vec{X})P(2)} = \frac{P(2|\vec{X})}{1-P(2|\vec{X})}\frac{P(1)}{P(2)}
\label{eq:bayes}
\end{equation}
We note that $P(1)/P(2) = N_1/N_2$, where $N_2$ and $N_1$ are the
number of samples from the two simulations respectively.  Although
either $P(\vec{X}|1)$ or $P(\vec{X}|2)$ can be eliminated, we are left
with one independent continuous free energy distribution. Writing
either $P(\vec{X}|2)$ or $P(\vec{X}|1)$ in a closed form is system
dependent; specifically, it depends on the unknown density of states.
We define the constant $M = \ln(N_F/N_R)$ and rewrite
Eq.~\ref{eq:ratio.mle} as:
\begin{eqnarray}
\frac{P(2|\vec{X})}{1-P(2|\vec{X})} = \exp(-M - \alpha_0 - \sum_{j=1}^M\alpha_j X_j)
\label{eq:revlog}
\end{eqnarray}
Given Eq.~\ref{eq:revlog}, we can rewrite the probability of a
single measurement $P(1|\vec{X}_i)$ or  $P(2|\vec{X}_i)$ as:
\begin{eqnarray}
P(1|\vec{X}_i) = \frac{1}{1+\exp(M+\vec{\alpha}\cdot\vec{X}_i)}\nonumber\\
P(2|\vec{X}_i) = \frac{1}{1+\exp(-M - \vec{\alpha}\cdot\vec{X}_i)}
\label{eq:logisitic}
\end{eqnarray}
The total likelihood of any given observation $X_i$ is the product of
all the individual likelihoods, giving:
\begin{eqnarray*}
\ln L (\vec{\alpha}|\mathrm{data}) &=& \sum_{i=1}^{N_1} \ln f(-M -  \vec{\alpha}\cdot\vec{X}_i) + \sum_{i=1}^{N_2} \ln f(M+\vec{\alpha}\cdot\vec{X}_i) 
\end{eqnarray*}
where $f(x) = [1+\exp(x)]^{-1}$ is the Fermi function.  This
likelihood equation can be minimized directly or by finding the
gradient with respect to the $\alpha$ parameters and solving for
$\nabla (\ln L) = 0$ to give the maximum likelihood result.  The
log likelihood function has a single minimum, and thus there will be
only a single root to $\nabla(\ln L)=0$.

The covariance matrix of each $\alpha_j$ can be written in terms of
the Fisher information: 
\begin{eqnarray}
\Var{\alpha_j} = I(\alpha_j)^{-1} = -\frac{1}{\frac{\partial^2 \ln L(\alpha)}{\partial \alpha_j^2}}.
\end{eqnarray}
Note that in Ref.~\cite{Shirts2003}, an additional factor dependent on
the number of samples was required to get the correct uncertainty
estimates.  In that case, we assumed that the simulation was conducted
properly, so that $\beta$ was known and thus had an additional
constraint, leaving only a single parameter estimated from ratio of
two distributions, which ends up reducing the uncertainty by this
constant factor.~\cite{Anderson1972} In this case, we are solving for
two parameters using the data from two distributions, and no implicit
constraints are applied.  Thus the correction is not required.

\end{appendix}

\bibliography{ShirtsJCTC2012}

\end{document}